\documentclass[journal]{IEEEtran}
\usepackage{amsmath}
\usepackage{amssymb}
\usepackage{amsfonts}
\usepackage{graphicx}
\usepackage{enumerate}
\usepackage{bbm}
\usepackage{subfig}
\usepackage{mathtools}
\usepackage{cite}
\usepackage{url}
\usepackage{array}
\usepackage{verbatim}
\usepackage{comment}
\usepackage{stfloats}
\usepackage{multirow}
\usepackage{hyperref}

\usepackage{hhline}
\newcommand*\dd{\mathop{}\!\mathrm{d}}

\newtheorem{theorem}{Theorem}
\newtheorem{lemma}{Lemma}

\newcounter{MYtempeqncnt}
\usepackage{color}
\definecolor{red}{rgb}{1,0,0}
\definecolor{blue}{rgb}{0,0,1}

\usepackage[font=small,skip=-1pt]{caption}
\IEEEoverridecommandlockouts
\begin{document}
\title{Angle Feedback for NOMA Transmission in mmWave Drone Networks}
%\title{Distance and Angle based Limited Feedback Schemes for NOMA transmission in mmWave Drone Networks}
\author{Nadisanka~Rupasinghe,~Yavuz~Yap{\i}c{\i},~\.{I}smail~G\"{u}ven\c{c},~Monisha~Ghosh~and~Yuichi~Kakishima
\thanks{N.~Rupasinghe,~Y.~Yap{\i}c{\i},~and~\.{I}.~G\"{u}ven\c{c} are with the Department of Electrical and Computer Engineering, North Carolina State University, Raleigh, NC, 27606 (e-mail:~\{rprupasi,yyapici,iguvenc\}@ncsu.edu).

Monisha Ghosh is with the Department of Computer Science, Uniersity of Chicago, Chicago, IL  (e-mail: monisha@uchicago.edu).

Yuichi Kakishima was with DOCOMO Innovations, Inc., Palo Alto, CA, 94304  (e-mail: yuichi.kakishima.vc@nttdocomo.com).

This research was supported in part by NSF under the grant CNS-1618692.
}}%
\maketitle

\begin{abstract}

In this paper, we consider an unmanned aerial vehicle (UAV) based wireless network using non-orthogonal multiple access (NOMA) transmission in millimeter-wave frequencies to deliver broadband data in a spectrally efficient fashion at hotspot scenarios. The necessity for the NOMA transmitter to gather information on user channel quality becomes a major drawback in practical deployments. We therefore consider various limited feedback schemes for NOMA transmission, to relieve the complexity of tracking and feeding back the full channel state information (CSI) of the users.
In particular, through beamforming we allow NOMA to exploit the space domain, and hence the user angle emerges as a promising (yet novel)
limited feedback scheme. We show that as the user region for NOMA transmission gets wider, the users become more distinctive at the transmitter side with respect to their angles, making user angle feedback a better alternative than distance feedback in such scenarios.
We rigorously derive and analyze the outage sum rate performance for NOMA transmission considering various user ordering strategies involving full CSI, angle, and distance feedback schemes. Our analytical results for NOMA outage sum rates using those feedback schemes match closely with simulations, and provide useful insights on properly choosing a limited feedback scheme for different deployment geometries and operating configurations.
\end{abstract}

\begin{IEEEkeywords}
5G, drone, HPPP, mmWave, non-orthogonal multiple access (NOMA), stadium, UAV.
\end{IEEEkeywords}

\section{Introduction}

% In \cite{Ding17PoorRandBeamforming}, a random beamforming approach for mmWave NOMA networks are proposed where a single BS is supporting users who are randomly distributed across a disk with respect to a homogeneous Poisson point process (HPPP). In that work, the BS generates a random beam which creates a wedge-shaped sector, and users within this wedge-shaped sector are then considered for NOMA transmission. In particular, two users with different effective channel gains are selected for NOMA transmission within each beam.
Unmanned aerial vehicles (UAVs) serving as aerial base stations (BSs) is emerging as a cost-effective and efficient solution for providing rapid on-demand connectivity during temporary events and after disasters \cite{IEEE_Spectrum, Zeng2016UAVOpportunities, Arvind2015UAV, UAV_NOMA_Asilomar}. There have been several recent use cases where UAVs were deployed as aerial BSs for providing temporary connectivity. For example, AT\&T had recently deployed their flying cell on wings (COW) to provide data, voice, and text services to users in Puerto Rico in the aftermath of hurricane Maria~\cite{Att2, Times}.
%Their telecommunication infrastructure had been destroyed by this devastating hurricane, and while they were restoring their network back, they had deployed flying COWs as flying BSs.
During this deployment, the UAV-BS was providing LTE connectivity to users in an area of up to $40$-square mile to restore the wireless network. %According to a recent article from BBC \cite{BBC},
Nokia and British mobile operator EE have been flying small quadcopter drone BSs in Scotland since 2016, %. Their main focus is
to provide instant LTE coverage using drone BSs over a disaster area with radius of about $31$~miles~\cite{BBC}. Further, AT\&T has been recently exploring the possibility of deploying UAV-BSs for augmenting their network capacity especially in hot spot scenarios, while Ericsson considers use of drone BSs as a viable solution to provide on-demand coverage in an area with bad signal specifically for music festivals~\cite{BBC}.

% This is exactly the scenario we are focusing on in this work.

%However,
Due to the limited energy resources on board of a UAV, achieving higher spectral efficiency (SE) is of paramount importance to reap maximum benefits from UAV based communication networks. In this regard, integrating non-orthogonal multiple access (NOMA) to UAV-BSs can be an effective solution to improve their SE  \cite{Docomo_NOMA}. In contrast to the conventional orthogonal multiple access (OMA) schemes (e.g., time-division multiple access (TDMA)), NOMA simultaneously serves multiple users in the same time, frequency, code or space resources in a non-orthogonal fashion by considering power domain for multiple access. Hence, UAV-BSs can serve multiple users simultaneously with NOMA using the same resources while enhancing the achievable SE. %  while enhancing SE. % specifically when the available resources are limited.

Use of NOMA techniques to improve SE has been studied extensively in the literature in a broader context. In particular, NOMA with multi-antenna transmission techniques has been recently receiving higher attention \cite{Ding16MIMO_NOMA,Ding16Schober_MIMO_NOMA,Ding17PoorRandBeamforming, Ding16MIMO_IoT,UAV_NOMA_Asilomar,NadisankaTCoM, Zeng17MIMO_NOMA_Capacity_Comp}. In \cite{Ding16MIMO_NOMA} multiple-input-multiple-output (MIMO) techniques are introduced to NOMA transmission along with user pairing and power allocation strategies to enhance MIMO-NOMA performance over MIMO-OMA. A general MIMO-NOMA framework applicable to both downlink (DL) and uplink (UL) transmission is proposed in \cite{Ding16Schober_MIMO_NOMA} by considering signal alignment concepts. A random beamforming approach for millimeter (mmWave) NOMA networks is proposed in \cite{Ding17PoorRandBeamforming}. In that, for user ordering, full channel state information (CSI) of users which depend on the angle offset between the randomly generated base station (BS) beam, user distances and small scale fading are considered. Two users are then served simultaneously within a single beam by employing NOMA. Recently, 3rd generation partnership project group (3GPP) has also been investigating to including NOMA in the next generation communication standards \cite{3GPP16NOMA_LTE,YChen_NOMA_Standardization}.

% NOMA transmission is introduced in \cite{UAV_NOMA_Asilomar} to serve multiple users within the same beam generated by a UAV-BS providing coverage in a stadium scenario. In that, effective channel gains are considered as the user ordering criteria for NOMA.

\subsection{Related work}
Use of NOMA to support UAV based communication networks has recently been explored in the literature. A UAV based mobile cloud computing system is proposed in \cite{7932157}, where UAVs, using NOMA transmission, offer computation offloading opportunities to mobile stations (MS) with limited local processing capabilities.
In \cite{Sharma17_UAV_NOMA}, without considering any multi-antenna techniques, two user NOMA transmission is introduced to fixed-wing type UAV acting as an aerial BS. In that, the UAV-BS moves in a circular trajectory around the center of a macrocell without altering its altitude. Power allocation and UAV altitude optimization approach is proposed in \cite{Sohail18_UAVAssistedCommun} to maximize achievable NOMA sum rates in a UAV based communication network. Max-min rate optimization problem is formulated in \cite{Nasir18_UAV_NOMA} for a scenario where single antenna UAV-BS is serving ground users employing NOMA transmission. In that, joint optimization of power, bandwidth, UAV altitude and antenna beamwidth is considered. A cooperative NOMA transmission approach is proposed in \cite{Nguyen18_CoopNOMA_UAV} for designing UAV-assisted wireless backhaul networks. In that, UAVs acting as aerial BSs provide coverage to distinct user clusters and during backhaul transmission from macro BS to UAVs, a cooperative NOMA strategy is introduced. In \cite{YLiu18_UAV_NOMA}, considering three case studies, performance of NOMA enabled UAV networks is investigated. In particular, a stochastic geometry based modeling for NOMA aided UAV networks is presented alongside joint power allocation and trajectory design for UAVs considering static NOMA users and machine learning based UAV placement approach when ground NOMA users are moving. A cyclical NOMA transmission strategy for UAV-enabled wireless networks is proposed in \cite{Sun18_CyclicalNOMA_UAV}. In that, the minimum throughput maximization over all ground users is considered by jointly optimizing multiuser communication scheduling with cyclical NOMA and UAV trajectory. In our earlier work \cite{UAV_NOMA_Asilomar, NadisankaTCoM}, NOMA is introduced to UAVs acting as aerial BSs to provide coverage over a stadium or a concert scenario. In particular, leveraging multi-antenna techniques a UAV-BS generates directional beams, and multiple users are served within the same beam employing NOMA transmission, i.e., reusing space resources. In \cite{UAV_NOMA_Asilomar} we assume the availability of full CSI feedback whereas in \cite{NadisankaTCoM} the availability of only user distance information is assumed as a practical feedback scheme for NOMA formulation. Note that, any of these prior works do not investigate explicit impact of the limited feedback schemes and user ordering strategies on NOMA transmission in UAV based communication networks. In \cite{UAV_NOMA_SPAWC}, solely based on the \emph{extensive computer simulations}, we have carried out an initial evaluation of the achievable performance with different limited feedback schemes and user ordering strategies for NOMA transmission in UAV based communication networks.

% In this paper, we consider a similar scenario as in \cite{UAV_NOMA_Asilomar,NadisankaTCoM_arXiv}, where a UAV-BS is employed to provide broadband connectivity over a densely packed user area in a stadium. Then,  NOMA transmission along with multi-antenna transmission is introduced to improve the SE. In order to relieve the burden owing to tracking and feeding back full CSI used by the NOMA transmitter for user scheduling and power allocation, we consider two limited feedback schemes as practical alternatives. In particular, 1) user distance (as in~\cite{NadisankaTCoM_arXiv}), and 2) user angle information, are considered as available user feedback which provide a measure of user channel quality. We show that, depending on the geometry of the area covered by UAV-BS, these limited feedback schemes can provide comparatively different sum rate performances. In particular, numerical results verify that angle as the limited feedback is significantly superior to distance feedback whenever users are more distinguishable by their respective angles (i.e., when the horizontal footprint of UAV-BS beam is sufficiently wider). Further, considering two user ordering criteria with angle feedback (Fej\'er kernel, and absolute angle) achievable sum rate performance under different geometries and NOMA user pairs are evaluated.

% In this paper, which is a significantly extended version of\textcolor{red}{~\cite{UAV_NOMA_SPAWC}}, we leverage the user angle information \textcolor{red}{as a promising candidate for the limited feedback mechanism to be employed in the mmWave NOMA communications.}

\subsection{Contributions}
In this paper, which is a significantly extended version of~\cite{UAV_NOMA_SPAWC}, we develop a rigorous analytical framework to evaluate achievable rate performance for NOMA transmission with different user ordering strategies considering the availability of different channel quality feedback. In particular, we assume a similar scenario as in \cite{UAV_NOMA_Asilomar, NadisankaTCoM} where a UAV-BS provides broadband coverage over a stadium/concert. Considering multi-antenna and mmWave transmission, UAV-BS generates directional beams and multiple users are served within a beam thanks to NOMA. Note here that, beamforming allows reusing space domain resources as well for NOMA in additional to time and frequency resources. Due to this reason, user angle becomes a promising practical feedback scheme for user ordering in NOMA formulations. There are some recent works in the literature looking into limited feedback based NOMA \cite{Yang16_NOMA_PArtialCSI, Ding17PoorRandBeamforming, NadisankaTCoM, Wan18_NOMA_PartialCSI}. However, to the best of the authors' knowledge, user angle as a feedback scheme for NOMA formulations  has not been studied in the literature before. Compared to the conventional limited feedback scheme based on users' distance (to the transmitter), the angle information is shown to have a significant potential in providing better separation for NOMA users in the power domain specifically for scenarios with multi-antenna transmission. In particular, the unique contributions of this work can be outlined as follows.  \begin{figure}[!t]
\begin{center}
\includegraphics[width=0.4\textwidth]{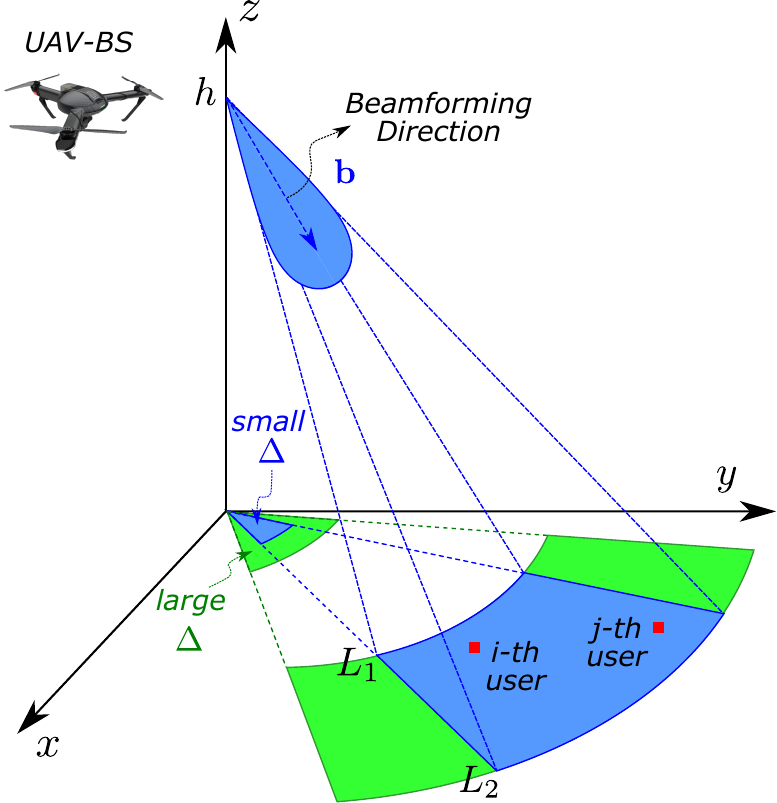}
\end{center}
\caption{System scenario of UAV-BS serving multiple users simultaneously in a single DL beam with NOMA transmission. The horizontal angle $\Delta$ is illustrated by relatively \textit{small} and \textit{large} angle values.}
\label{fig:footprint}
\vspace{0.25 em}
\end{figure}

\begin{itemize}
 \item[-] We rigorously analyze the outage probability and sum rate performance of the user ordering strategies under consideration. In particular, we propose a unified expression for the conditional outage probability, where the contribution of the ordered distance, (absolute) angle, and Fej\'er kernel distributions are explicitly provided. The respective ordered distributions are derived rigorously, as well, employing stochastic geometry and order statistics.
\item[-] We introduce user angle as a limited feedback scheme for NOMA which is an effective alternative for the full CSI and distance feedback schemes. Using the developed analytical framework, we show that the optimal use of the angle information while ordering users is to employ the Fej\'er kernel function. This is because, Fej\'er kernel fully captures the contribution of the user angle within the effective channel gain. Furthermore, absolute angle is also proposed as a more insightful yet suboptimal alternative to Fej\'er kernel based ordering.
 \item[-] We compare the performance of distance and angle based limited feedback and ordering strategies for various user region geometries. We show that as the users become more \textit{distinctive} based on their angle information (i.e., wider user regions), the angle feedback scheme and respective ordering strategies (i.e., absolute angle and Fejer kernel based ordering) significantly outperforms the classical distance based feedback and ordering.
\end{itemize}

The rest of the paper is organized as follows. Section~\ref{sec:Sys_Model} presents the system model while limited feedback schemes and respective user ordering strategies for NOMA are discussed in Section~\ref{Sec:NOMA_Transmission}. In Section~\ref{sec:Analytical_Rate_derivation} NOMA outage sum rates are analyzed and respective numerical results are presented in Section~\ref{sec:Numerical_results}. Finally, Section~\ref{sec:conclusion} provides concluding remarks.

%\textit{Notations:} $(\cdot)^{\rm H}$ and $(\cdot)^{\rm T}$ stand for the Hermitian and transpose operations. ${\rm sgn}(\cdot)$ denotes the signum function. $\mathcal{CN}(a,b)$ denotes complex Gaussian distribution with mean $a$ and variance $b$.

\section{System Model} \label{sec:Sys_Model}

We consider a mmWave-NOMA transmission scenario where a single UAV-BS equipped with an $M$~element uniform linear array (ULA), which is placed horizontally, is serving $K$ single-antenna users in the DL. We assume that all these users lie inside a specific \emph{user region}, which is identified by the inner-radius $L_1$, outer-radius $L_2$, and angle $\Delta$ as shown in Fig.~\ref{fig:footprint}. Specifically, $\Delta$ stands for the fixed angle within the projection of the horizontal beamwidth of the antenna pattern on the $xy$-plane. We assume that the UAV-BS generates a 3-dimensional (3D) beam $\textbf{b}$ as shown in Fig.~\ref{fig:footprint}, which is adjusted in the horizontal and vertical domains adequately (i.e., electronically and mechanically, respectively) to cover the user region entirely. All users are represented by the index set $\mathcal{N}_{\rm U} = \{1,2,\ldots, K\}$ with the cardinality $|\mathcal{N}_{\rm U}|$ being equal to $K$. Note that it is possible to model various different hot spot scenarios reasonably (e.g., stadium, concert hall, traffic jam, and urban canyon, etc.) by modifying these parameters. Importantly, these hot spot scenarios are of high practical importance as discussed in \cite{BBC}. Further, we assume UAV stability and orientation are well maintained through some advanced techniques \cite{DroneOrientation_1,DroneOrientation_2,DroneOrientation_3}, ensuring there is no impact due to orientation drifts on UAV-BS DL transmission.

We consider that the users are randomly distributed following a homogeneous Poisson point process (HPPP) with density $\lambda$. Hence, the number of users in the specified user region is Poisson distributed such that $\textrm{P}(K \textrm{ users in the user region})\,{=}\, \frac{\mu^K e^{{-}\mu}}{K!} $ with $\mu\,{=}\,(L_2^2\,{-}\,L_1^2)\frac{\Delta}{2} \lambda$.

The channel $\textbf{h}_k$ between the UAV-BS and user $k$ is
\begin{align} \label{k_UE_original_channel}
\textbf{h}_k = \sqrt{M} \sum \limits _{p=1}^{N_{\rm P}} \frac{\alpha_{k,p} \, \textbf{a}(\theta_{k,p})}{\sqrt{\textrm{PL}\big(\sqrt{d_k^2 + h^2}\,\big)}},
\end{align} where $N_{\rm P}$, $h$, $d_k$, $\alpha_{k,p}$ and $\theta_{k,p}$ represent the number of multi-paths, UAV-BS hovering altitude, horizontal distance between user $k$ and UAV-BS, gain of the $p$-th path which follows standard complex Gaussian distribution with $\mathcal{CN}(0,1)$, and angle-of-departure (AoD) of the $p$-th path, respectively. The steering vector $\textbf{a}(\theta_{k,p})$ corresponding to the AoD $\theta_{k,p}$ is defined as \vspace{-1em}

\small
\begin{align*}
\textbf{a}\left( \theta_{k,p} \right) = \frac{1}{\sqrt{M}} \left[ 1 \;\; e^{-j2\pi \frac{D}{\zeta}\sin\left( \theta_{k,p}\right) } \; \dots \; e^{-j2\pi \frac{D}{\zeta}\sin\left( \theta_{k,p} \right)\left( M-1\right) } \right]^{\rm T},
\end{align*}\normalsize
where $D$ is the antenna spacing of ULA and $\zeta$ is the wavelength. The path loss (PL) between user $k$ and UAV-BS is captured by $\textrm{PL}\left(\sqrt{d_k^2 + h^2}\right)$. Without any loss of generality, we assume that all the users have line-of-sight (LoS) paths since UAV-BS is hovering at relatively high altitudes and the probability of having scatterers around UAV-BS is very small. In particular, as presented in \cite{ mmWave_Channel_Ch_Modeling_2, mmWave_Channel_Ch_Modeling_1}, the gains of Non-LoS (NLoS) paths are typically $20$~dB weaker than LoS path in mmWave channels. Hence, as considered in \cite{Lee16mmWaveChannel, mmWave_Channel_Cap_Analysis}, it is reasonable to assume only LoS path for the mmWave channel under consideration, and \eqref{k_UE_original_channel} accordingly becomes
\begin{align} \label{k_UE_modified_LoS_channel}
\textbf{h}_k = \sqrt{M} \frac{\alpha_k \, \textbf{a}(\theta_k)}{\sqrt{\textrm{PL}\big(\sqrt{d_k^2 + h^2}\,\big)}} \,,
\end{align}
where $\alpha_{k}$ and $\theta_{k}$ are the complex gain and AoD of the LoS path, respectively.

\section{NOMA Transmission with Limited Feedback} \label{Sec:NOMA_Transmission}
In this section, we consider outage sum rate formulations and user ordering strategies for the NOMA transmission, which will be used within the respective derivations later in Section~\ref{sec:Analytical_Rate_derivation}. We assume that each user has its own QoS based target rate. Further, we consider each user sends limited information back to the NOMA transmitter on its channel quality, which involves either distance or angle information of that particular user.

\subsection{Outage Probabilities and Sum Rates} \label{sec:Outage_Prob_Sum_Rates}

We assume that the single UAV-BS may be assigned to either the entire environment where users are distributed, e.g., a stadium, or a part of it, e.g., a sector of a stadium. The AoD $\overline{\theta}$ of the beam $\textbf{b}$ generated by UAV-BS is therefore assumed to take values either from $[0{,}\,2\pi]$, or a subset of it. In addition, the full coverage of the entire environment can be performed by choosing values for $\overline{\theta}$ from its support either sequentially or randomly over time. Without any loss of generality, the effective channel gain of user $k\,{\in}\,\mathcal{N}_\textrm{U}$ for a beamforming direction $\overline{\theta}$ of UAV beam $\textbf{b}$ is given using (2) as follows
\begin{align} \label{eq:Eff_channel_gain_original}
|\textbf{h}_k^{\rm H}\textbf{b}|^2 &= \frac{|\alpha_k|^2 |\textbf{b}^{\rm H}\textbf{a}(\theta_k)|^2}{M  \times \textrm{PL}\left(\sqrt{d_k^2 + h^2}\right)}
\nonumber \\
&= \frac{|\alpha_k|^2 M}{\textrm{PL}\left(\sqrt{d_k^2 + h^2}\right)}
 \left| \frac{ \sin \left( \frac{\pi M(\sin \overline{\theta}-\sin \theta_k)}{2} \right)}{  M\sin \left( \frac{\pi (\sin \overline{\theta}-\sin \theta_k)}{2} \right)}\right|^2 \,
\end{align}
where we assume a critically spaced ULA, i.e. $D = \frac{\zeta}{2}$. Following the convention of~\cite{Ding17PoorRandBeamforming, NadisankaTCoM}, we assume small $2\Delta$ while analyzing sum rates, i.e., $2\Delta\,{\rightarrow}\,0$, which results in small angular offset such that $|\overline{\theta}\,{-}\,\theta_k|\,{\rightarrow}\,0$. In particular, for mmWave transmission, small $\Delta$ is a reasonable assumption. Choosing the coordinate system appropriately, small angular offset implies small individual angles such that $\sin\overline{\theta}\,{\rightarrow}\,\overline{\theta}$ and $\sin\theta_k\,{\rightarrow}\,\theta_k$, and \eqref{eq:Eff_channel_gain_original} can be approximated as
\begin{align} \label{eq:Eff_channel_gain}
|\textbf{h}_k^{\rm H}\textbf{b}|^2 &\approx \frac{|\alpha_k|^2}{M \times\textrm{PL}\left(\sqrt{d_k^2 + h^2}\right)}
 \left| \frac{ \sin \left( \frac{\pi M(\overline{\theta} - \theta_k)}{2} \right)}{  \sin \left( \frac{\pi (\overline{\theta} - \theta_k)}{2} \right)}\right|^2
 \nonumber \\
 &= \frac{|\alpha_k|^2}{\textrm{PL}\left(\sqrt{d_k^2 + h^2}\right)}  \tilde{{\rm F}}_M(\pi [\overline{\theta} - \theta_k]),
\end{align}
where $\tilde{{\rm F}}_M(\cdot)$ is called Fej\'er kernel. For notational simplicity, we define ${\rm F}_M(\overline{\theta}-\theta_k)=\tilde{\rm F}_M(\pi[\overline{\theta}-\theta_k])$ and use this modified notation through the rest of the paper.

%We denote the azimuth AoD value associated with the DL beam $\textbf{b}$ (see Fig.~\ref{fig:footprint}) as $\overline{\theta}$, which might be considered as \textit{beamforming direction} for this setting, as well. Assuming that the horizontal angle $\Delta$ is small as compared to AoD values for user $k$ (i.e., $\theta_k$) or the DL beam (i.e., $\overline{\theta}$), we have a small angular offset such that $|\overline{\theta}\,{-}\,\theta_k|\,{\rightarrow}\,0$. Note that in an adequate coordinate system, small angular offset always implies small individual angles such that $\sin\overline{\theta}\,{\rightarrow}\,\overline{\theta}$ and $\sin\theta_k\,{\rightarrow}\,\theta_k$. Hence, as shown in \cite{NadisankaTCoM}, using \eqref{k_UE_modified_LoS_channel}, the effective channel gain of user $k$ for a particular beamforming direction $\overline{\theta}$ is given by
%\begin{align} \label{eq:Eff_channel_gain}
%|\textbf{h}_k^{\rm H}\textbf{b}|^2 &\approx \frac{|\alpha_k|^2}{M \times\textrm{PL}\left(\sqrt{d_k^2 + h^2}\right)}
% \left| \frac{ \sin \left( \frac{\pi M(\overline{\theta} - \theta_k)}{2} \right)}{  \sin \left( \frac{\pi (\overline{\theta} - \theta_k)}{2} \right)}\right|^2 \nonumber \\
%&= \frac{|\alpha_k|^2}{\textrm{PL}\left(\sqrt{d_k^2 + h^2}\right)} {\rm F}_M(\overline{\theta} - \theta_k),
%\end{align}
%where ${\rm F}_M(\cdot)$ is the well-known Fej\'er kernel function.

We assume without any loss of generality that the users in set $\mathcal{N}_{\rm U}$ are already indexed from the best to the worst channel quality under a given criterion, where the details of ordering strategies considered in this work will be substantiated in the next section. In addition, we also assume that the NOMA transmission targets to serve $K_{\rm N}$ users simultaneously with $K_{\rm N}\,{\leq}\,K$, and that $\mathcal{N}_{\rm N}$ denotes the indices of NOMA users obeying the original order of $\mathcal{N}_{\rm U}$ and $\mathcal{N}_{\rm N} \,{\subset}\, \mathcal{N}_{\rm U}$. It is worth remarking here that, since UAV-BS employs NOMA transmission, multiple users can be served within the same DL beam. Compared to the conventional beamforming, this way the achievable SE can be further enhanced. %In the following, NOMA users are chosen  such that $\mathcal{N}_{\rm N} \,{=}\, \{1,2,\ldots, K_{\rm N}\}$ for ease of representation, although the the results are valid for non-sequential index sets, as well.

Defining $\beta_k$ as the $k$-th user power allocation coefficient, with $\beta_j\,{\leq}\,\beta_i$ for $\forall \, i\,{\geq}\,j$, $i,j \,{\in}\, \mathcal{N}_{\rm N}$, and $\sum_{k\in \mathcal{N}_{\rm N}} \beta_k^2\,{=}\,1$. The transmitted signal $\textbf{x}$ is generated by superposition coding as
\begin{align}
\textbf{x} = \sqrt{P_{\rm Tx}}\textbf{b}\sum \limits_{k\in \mathcal{N}_{\rm N}} \beta_k s_{k},
\end{align}
where $P_{\rm Tx}$ and $s_{k}$ with $\mathbb{E}\left( \left| s_k \right|^2 \right) = 1$ are the total DL transmit power and $k$-th user's message, respectively. The received signal at the $k$-th user is then given as
\begin{align} \label{eq:k-th_user_Rx_signal}
y_{k}= \textbf{h}_{k}^{\rm H} \textbf{x} +  v_k = \sqrt{P_{\rm Tx}}\textbf{h}_{k}^{\rm H} \textbf{b}\sum \limits_{k\in \mathcal{N}_{\rm N}} \beta_k s_{k} + v_k,
\end{align}
where $v_k$ is a zero-mean complex Gaussian noise with variance $N_0$ denoted as $\mathcal{CN}(0,N_0)$.

At the receiver, each user first decodes messages of all weaker users (allocated with larger power) sequentially in the presence of stronger users’ messages (allocated with smaller power), provided that required QoS based target rates of weaker users are feasible and all met. Those decoded messages are then subtracted from the received signal in \eqref{eq:k-th_user_Rx_signal}, and the user decodes its own message treating the stronger users' messages as noise. This overall decoding process is known as successive interference cancellation (SIC).

When the instantaneous rate while decoding any of the weaker users' message is larger than the QoS based target rate of that weak user, the associated decoding operation occurs without any error. Assuming that all interfering messages of users weaker than $m$-th user are decoded accurately (without any error), $k$-th user has the following signal-to-interference-plus-noise ratio (SINR) while decoding $m$-th user message
\begin{align} \label{eq:SINR_mk_th_user}
\textrm{SINR}_{m{\rightarrow}k} = \frac{P_{\rm Tx}|\textbf{h}_{k}^{\rm H}\textbf{b}|^2 \beta_{m}^2}{P_{\rm Tx} \sum \limits_{l < m,\, l \in \mathcal{N}_{\rm N}}|\textbf{h}_{k}^{\rm H}\textbf{b}|^2 \beta_{l}^2 + N_0},
\end{align}
where $m\,{\geq}\,k$. Actually, \eqref{eq:SINR_mk_th_user} also represents the SINR of $k$-th user while decoding its own message for $m\,{=}\,k$. Note that the summation in the denominator of \eqref{eq:SINR_mk_th_user} disappears while $k$-th user being the strongest one is decoding its own message (i.e., no possible $l$ index satisfying $l \,{<}\, k$ for this specific case), and we have $\frac{P_{\rm Tx}}{N_0}|\textbf{h}_{k}^{\rm H}\textbf{b}|^2 \beta_{k}^2$.

As captured by the SINR in \eqref{eq:SINR_mk_th_user}, $k$-th ($k\in \mathcal{N}_{\rm N} $) user having a better channel quality than $m$-th ($m\in \mathcal{N}_{\rm N} $) user, decodes $m$-th user message. Defining the instantaneous rate associated with $k$-th user decoding $m$-th user message as $R_{m{\rightarrow}k}\,{=}\,\log_2 \left( 1\,{+}\,\textrm{SINR}_{m{\rightarrow}k}\right)$, the conditional outage probability of the $k$-th user can be given as
\begin{align}
\textrm{P}_{k|\mathcal{S}_{K}}^{o}  &= 1 - \textrm{P} \left( R_{K_{\rm N}{\rightarrow}k} > \overline{R}_{K_{\rm N}}, \cdots ,R_{k} > \overline{R}_k | \, \mathcal{S}_K \right) \label{eq:Outage_k_th_user_1}\\
&= 1 - \textrm{Pr} \left( \bigcap\limits_{l \geq k,\, l \in \mathcal{N}_{\rm N}} R_{l{\rightarrow}k} > \overline{R}_{l} \, \Big| \, \mathcal{S}_{K} \right) \label{eq:Outage_k_th_user_2} \\
&= 1 - \textrm{Pr} \left( \bigcap\limits_{l \geq k,\, l \in \mathcal{N}_{\rm N}} \textrm{SINR}_{l{\rightarrow}k} > \epsilon_{l}  \, \Big| \,  \mathcal{S}_{K} \right), \label{eq:Outage_k_th_user_3}
\end{align}

\noindent where $\overline{R}_k$ is the QoS based target rate for $k$-th user and  $\epsilon_k\,{=}\,2^{\overline{R}_k}\,{-}\,1$. As represented in \eqref{eq:Outage_k_th_user_1}, $k$-th user can successfully decode his message only if he can decode all the messages of users having worst channel qualities than him, i.e.,  $R_{K_{\rm N}{\rightarrow}k} > \overline{R}_{K_{\rm N}}, \cdots ,R_{k} > \overline{R}_k $. A compressed version of \eqref{eq:Outage_k_th_user_1} is captured in \eqref{eq:Outage_k_th_user_2} considering set notations. Rather than representing conditional outage in rate terms, \eqref{eq:Outage_k_th_user_3} represents that in SINR terms.

Note that \eqref{eq:Outage_k_th_user_2} is \textit{conditioned} over $\mathcal{S}_{K}$, which describes the given condition on number of user, $K$ in the user region. This conditioning is important for deriving outage probabilities analytically under different user ordering strategies as captured in Section~\ref{sec:Analytical_Rate_derivation}. More specifically, $\mathcal{S}_{K}$ might represent either a particular integer \textit{one at a time}, i.e., $\mathcal{S}_{K_1}{:} \left\lbrace K \,|\, K\,{=}\,i \right\rbrace$, or a range of integers \textit{jointly}, i.e., $\mathcal{S}_{K_2}{:}\left\lbrace K \,|\, j\,{\leq}\, K \,{<}\, i \right\rbrace$, where $i,\, j \in \mathbb{Z}^+$. In other words, the outage probability in \eqref{eq:Outage_k_th_user_2} corresponds to a single $K$ value for $\mathcal{S}_{K_1}$, while it is associated with a range of $K$ values jointly for $\mathcal{S}_{K_2}$.

When the outage probability in \eqref{eq:Outage_k_th_user_2} is obtained using $\mathcal{S}_{K_1}$, the outage sum rate is given as
\begin{align} \label{eq:sum_rate_NOMA_singleK}
&R^{\textrm{NOMA}} = \textrm{Pr} \left( K{=}1 \right) \big(1{-} \tilde{\textrm{P}}_{1|\mathcal{S}_{K_1}}^{o} \big) \overline{R}_1 \nonumber \\
&+ \sum\limits_{n=2}^{\infty} \textrm{Pr} \left( K{=}n \right) \Bigg( \sum\limits_{\substack{k \leq n, \, k \in \mathcal{N}_{\rm N}}} \big( 1{-}\textrm{P}_{k|\mathcal{S}_{K_1}}^{o} \big) \overline{R}_k \Bigg),
\end{align}
where  $\tilde{\textrm{P}}_{k|\mathcal{S}_{K_{\tau}}}^{o}\,{=}\,\textrm{Pr}\left(\frac{1}{K_{\rm N}} \log\left(1{+}P_{\rm Tx}|\textbf{h}_{k}^{\rm H}\textbf{b}|^2/N_0 \right)\,{<}\,\overline{R}_k |\mathcal{S}_{K_{\tau}}\right)$ with $\tau\,{\geq}\, 1$ is the outage probability of $k$-th user during OMA transmission with the factor $\frac{1}{K_{\rm N}}$ capturing the associated loss in degrees-of-freedom (DoF) gain. For comparison, we consider OMA sum rates being computed in the same way as \eqref{eq:sum_rate_NOMA_singleK}, except that $\textrm{P}_{k|\mathcal{S}_{K_1}}^{o}$ in the inner summation is replaced with $\tilde{\textrm{P}}_{k|\mathcal{S}_{K_1}}^{o}$.

On the other hand, whenever \eqref{eq:Outage_k_th_user_2} is obtained for a set of $\mathcal{S}_{K_{\tau}}$'s with $\tau\,{\geq}\, 2$, where each $\mathcal{S}_{K_{\tau}}$ has a unique range of integers, the combined outage sum rate is described as \vspace{-0.5em}

\small
\begin{align} \label{eq:sum_rate_NOMA_setK}
R^{\textrm{NOMA}} &{=} \sum \limits_{\tau\geq 2} {\rm Pr} \left\lbrace \mathcal{S}_{K_{\tau}} \right\rbrace \sum \limits_{k \in \mathcal{N}_{\rm N}}\big(1{-}\textrm{P}_{k|\mathcal{S}_{K_{\tau}}}^{o}\big) \overline{R}_k{=}\sum\limits_{k \in \mathcal{N}_{\rm N}} \big(1{-} \textrm{P}_{k}^{o}\big) \overline{R}_k,
\end{align} \normalsize
where $\textrm{P}_{k}^{o}$ is called the \textit{unconditional} outage probability of $k$-th $(k\,{\in}\,\mathcal{N}_{\rm N})$ user. Note that whenever we have $K_{\rm N}\,{=}\,1$, single user transmission is employed with full time-frequency resources and transmit power are allocated to the single user scheduled. In addition, corresponding sum rates of OMA transmission can be computed using \eqref{eq:sum_rate_NOMA_setK} by replacing $\textrm{P}_{k|\mathcal{S}_{K_{\tau}}}^{o}$ with $\tilde{\textrm{P}}_{k|\mathcal{S}_{K_{\tau}}}^{o}$.

\subsection{Limited Feedback Schemes and User Ordering Strategies} \label{sec:feedback_noma}

Since NOMA transmitter allocates power to its users based on their channel qualities, it needs to order users according to their effective channel gains given in \eqref{eq:Eff_channel_gain}. This strategy therefore requires users to send appropriate information on their respective channel qualities back to the transmitter. When the underlying channel experiences rapid fluctuations over time, tracking of the full CSI at user terminals becomes cumbersome, and frequently sending this information back to the transmitter increases link overhead. Thus, we consider two types of \emph{limited feedback} schemes based on the 1) user distance $d_k$, and 2) user angle ${\theta}_k$. Both distance $d_k$ and angle $\theta_k$ information change much slowly as compared to full CSI, and are, hence, practical alternatives to full CSI feedback.
%Hence, users have to feedback either their distance or angle information to the UAV-BS and user ordering for NOMA formulation is done considering this feedback information.

Note that the user distance $d_k$ and angle $\theta_k$ appear in the effective channel gain expression of \eqref{eq:Eff_channel_gain} within the individual terms ${\textrm{PL}(\sqrt{d_k^2 \,{+}\, h^2})}$ and ${\rm F}_M(\overline{\theta} \,{-}\, \theta_k)$, respectively. While ${\textrm{PL}}(\cdot)$ is a monotonic function of $d_k$, ${\rm F}_M(\cdot)$ is not monotonically varying with $\theta_k$. Hence, distance $d_k$ is equivalent to ${\textrm{PL}(\sqrt{d_k^2 \,{+}\, h^2})}$ based user ordering whereas $\theta_k$ is not equivalent to ${\rm F}_M(\overline{\theta} \,{-}\, \theta_k)$ for ordering purpose. We therefore consider the following optimal ordering strategies, which are based on the available limited feedback information (being the distance or angle) as follows
\begin{align}
\textrm{Distance:} & \quad d_1 \leq d_2 \leq \dots \leq d_K \, ,\label{eq:distance_ordering} \\
\textrm{Fej\'er Kernel:} & \quad {\rm F}_M(\overline{\theta} \,{-}\, \theta_1) \geq \dots \geq {\rm F}_M(\overline{\theta} \,{-}\, \theta_K) \, ,\label{eq:Fejer_ordering}
\end{align}
where both these ordering strategies guarantee user ordering from the best to the worst channel quality, and therefore align with the formulations in Section~\ref{sec:Outage_Prob_Sum_Rates}. Although ${\rm F}_M(\cdot)$ is not a monotonic function of $\theta_k$, we will also consider the following suboptimal ordering scheme
\begin{align}
\textrm{Angle: } \quad \tilde{\theta}_1 \leq \tilde{\theta}_2 \leq \dots \leq \tilde{\theta}_K \,, \quad\quad\quad\quad\quad \label{eq:angle_ordering}
\end{align}
where $\tilde{\theta}_k$ is the absolute angle defined as $\tilde{\theta}_k\,{=}\,|\bar{\theta}{-}\theta_k|$. Note that, the general trend in \eqref{eq:angle_ordering} targets the ordering of users from the best to the worst channel quality. However, the channel qualities corresponding to \eqref{eq:angle_ordering} do not necessarily follow a strictly best-to-worst order since ${\rm F}_M(\overline{\theta} \,{-}\, \theta_k)$ in \eqref{eq:Eff_channel_gain} is not monotonic in $\theta_k$. Nevertheless, angle based ordering strategy of \eqref{eq:angle_ordering} generally gives more insight than optimal ordering of complicated function ${\rm F}_M(\overline{\theta} \,{-}\, \theta_k)$ as given in \eqref{eq:Fejer_ordering}, and produces the same performance under certain circumstances which will be detailed in Section~\ref{sec:Numerical_results}.

%Based on the two types of limited feedback schemes, three different criteria can be identified from \eqref{eq:Eff_channel_gain} for ordering users during NOMA formulation. They are, 1) distance, 2) Fej\'er kernel, and 3) absolute angle defined as $\tilde{\theta}_k\,{=}\,|\bar{\theta}{-}\theta_k|, \, k \in \mathcal{N}_{\rm U}$ based user ordering. In order to align with the derived outage probability and sum rate expressions in Section~\ref{sec:Outage_Prob_Sum_Rates}, user ordering with respect to above criteria considering best to worst channel quality can be given as follows. \begin{gather}
% \textrm{Distance: } d_1 \leq d_2 \leq\dots \leq d_K \label{eq:distance_ordering} \\
% \textrm{Fej\'er kernel: } {\rm F}_M(\theta_1) \geq {\rm F}_M(\theta_2) \geq\dots \geq {\rm F}_M(\theta_K) \label{eq:Fejer_ordering} \\
% \textrm{Angle: } \tilde{\theta}_1 \leq \tilde{\theta}_2 \leq\dots \leq \tilde{\theta}_K \label{eq:angle_ordering}
% \end{gather} where for notational simplicity we drop $\bar{\theta}$ in Fej\'er kernel term in \eqref{eq:Eff_channel_gain}. In the next section, we evaluate achievable rates with these feedback schemes and ordering criteria under different geometries considering $i$-th and $j$-th users only, though results can be generalized to multiple NOMA users as well.

\subsection{Effect of User Ordering on Angle/Distance Distributions}\label{sec:effect_ordering}

In this section, we study the impact of various user ordering strategies considered in Section~\ref{sec:feedback_noma} on the distributions of user angle and distance information together with Fej\'er Kernel function (as a function of the angle). The conclusion of the discussion herein will be used later in Section~\ref{sec:Analytical_Rate_derivation} while deriving exact outage sum rates for the NOMA transmission. We first consider the statistical relation between angle and distance of an arbitrary user in the following Lemma.

\begin{lemma} \label{the:unordered_ang_dist}
%\textit{\textcolor{red}{Lemma:}}
The distance and angle of an arbitrary user are statistically independent of each other given the user region and deployment scheme defined in Section~\ref{sec:Sys_Model}.
\end{lemma}
\begin{IEEEproof}
Consider the joint CDF of the distance $d_k$ and angle $\theta_k$ of user $k$ given as
\begin{align}\label{eqn:indep_1}
{\rm F}_{d_k\theta_k}(x,y) & = {\rm Pr} \left\lbrace d_k \leq x, \theta_k \leq y \right\rbrace \nonumber \\
&= \int_{0}^{y} {\rm Pr} \left\lbrace d_k \leq x \mid \theta_k = z \right\rbrace f_{\theta_k}(z) \dd z ,
\end{align}
which follows directly from basic probability theorems in~\cite{RossIntProbMod}. Recalling that users are uniformly deployed following HPPP, the probability within the integral of \eqref{eqn:indep_1} does not depend on the instantaneous angle value $z$. Note that, we can geometrically interpret in Fig.~\ref{fig:sketch_user_region}\subref{fig:user_region_angle} that any particular $\theta_k$ value does not alter either range or distribution of the distance $d_k$. As a result, \eqref{eqn:indep_1} can be manipulated as
\begin{align}\label{eqn:indep_2}
{\rm F}_{d_k\theta_k}(x,y) & = {\rm Pr} \left\lbrace d_k \leq x \right\rbrace \int_{0}^{y} f_{\theta_k}(z) \dd z \nonumber \\
&= {\rm Pr} \left\lbrace d_k \leq x \right\rbrace {\rm Pr} \left\lbrace \theta_k \leq y \right\rbrace ,
\end{align}
which shows the independence of distance $d_k$ and angle $\theta_k$, and hence completes the proof.
\end{IEEEproof}

\begin{figure}[!h]
\vspace{-0.2in}
\centering
\captionsetup[subfigure]{oneside,margin={0.5cm,0.5cm}}
\subfloat[PDF of \textit{ordered} distance for $j$-th user.]{\includegraphics[width=0.25\textwidth]{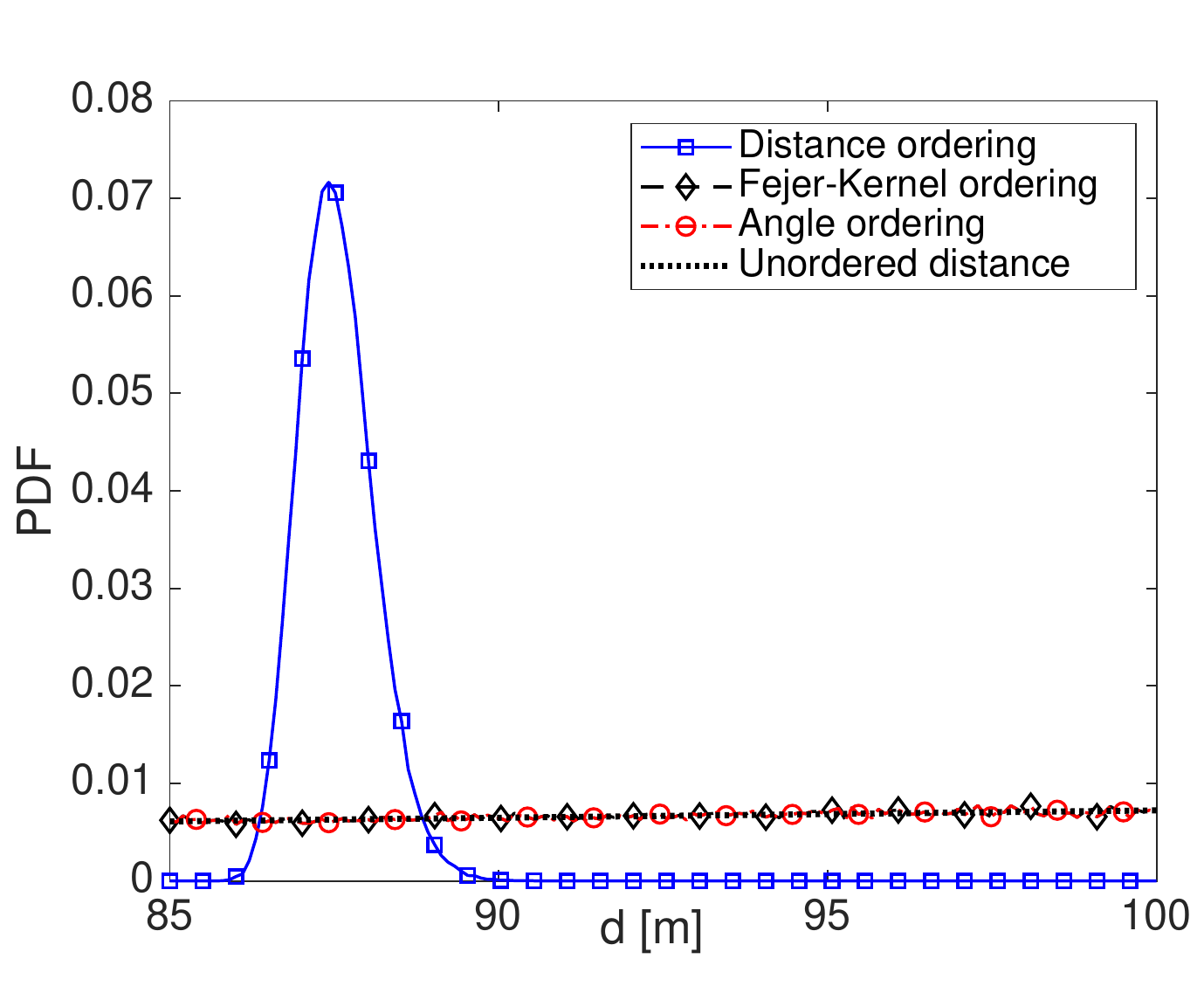}
\label{fig:distance_dist}}
\subfloat[PDF of \textit{ordered} angle for $j$-th user.]{\includegraphics[width=0.25\textwidth]{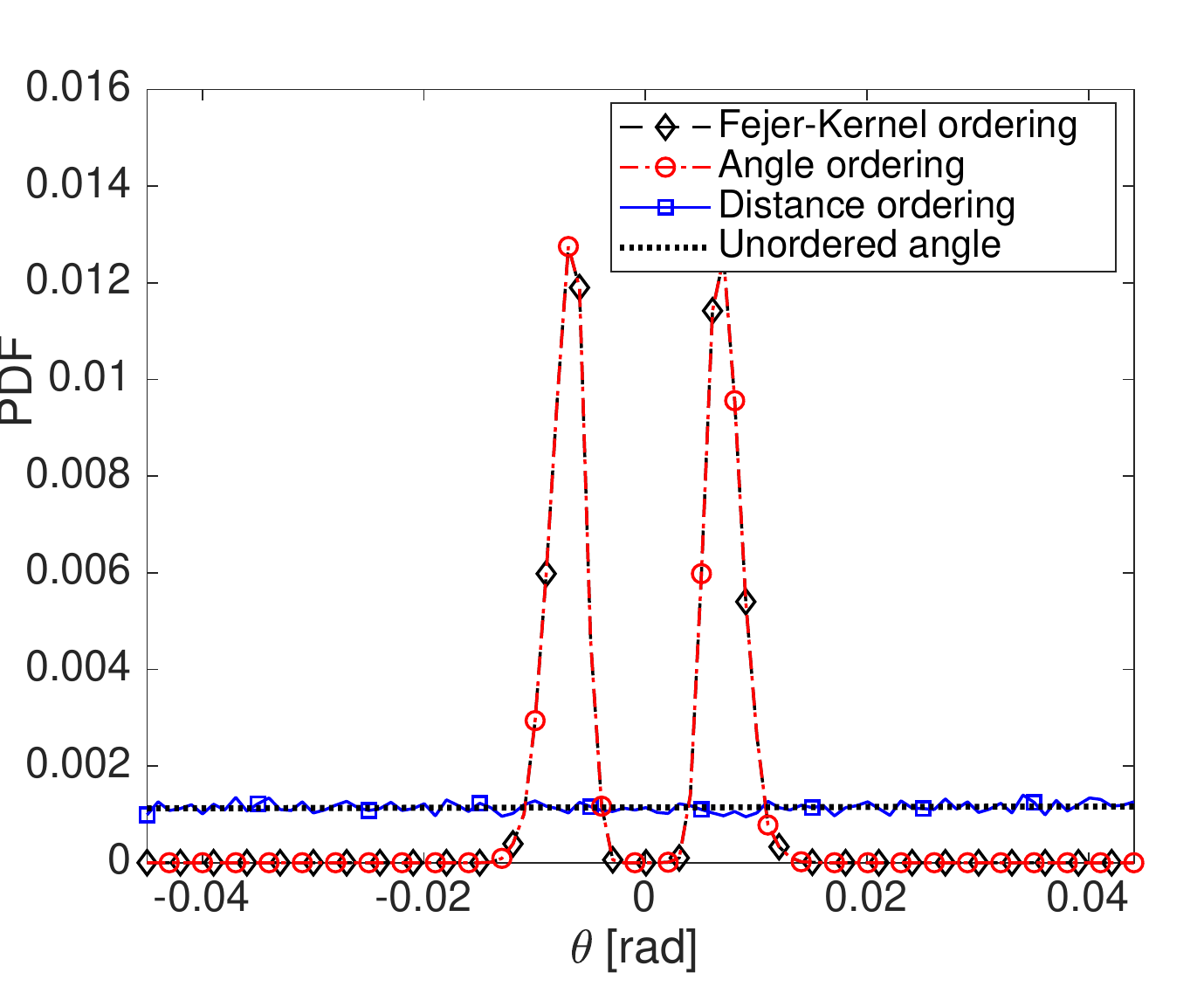}
\label{fig:angle_dist}}
\vspace{0.05in}
\caption{PDF of \textit{ordered} distance and angle of $j$-th user for the limited feedback based ordering strategies of Section~\ref{sec:feedback_noma} with $\Delta\,{=}\,5^{\circ}$ and $j\,{=}\,20$.}
\label{fig:angle_distance_dist}
\end{figure}

We now consider the impact of ordering strategies of Section~\ref{sec:feedback_noma} on the distance and angle distributions, where the ordering is performed based on the type of limited feedback information (either distance or angle). In Fig.~\ref{fig:angle_distance_dist}, we depict the PDF of the distance and angle of the $j$-th user associated with the ordering of users based on distance, Fej\'er Kernel, and angle given by \eqref{eq:distance_ordering}, \eqref{eq:Fejer_ordering}, and \eqref{eq:angle_ordering}, respectively. We observe in Fig.~\ref{fig:angle_distance_dist}\subref{fig:distance_dist} that the PDF of the ordered distance follows unordered distribution whenever the ordering strategy depends solely on angle (i.e., Fej\'er Kernel and angle ordering), and alters if distance ordering strategy is utilized. Similarly, the PDF of the angle in Fig.~\ref{fig:angle_distance_dist}\subref{fig:angle_dist} does not change when the distance ordering strategy is employed, and alters whenever one of the angle based ordering strategies (i.e., Fej\'er Kernel and angle ordering) is employed. Note that, although we do not show explicitly due to space limitations, the above conclusion for the angle distribution also applies to Fej\'er Kernel distribution, as expected. % In the next section, we will use both Theorem~\ref{the:unordered_ang_dist} and these results while deriving exact outage probabilities.
% \begin{figure}[!t]
% \vspace{-1em}
% \centering
% \hspace*{-0.3in}
% \subfloat[PDF of \textit{ordered} distance for $j$-th user]{\includegraphics[width=0.25\textwidth]{./img/Distance_Distirbutions}
% \label{fig:distance_dist}}
% %\hspace*{-0.2in}
% \subfloat[PDF of \textit{ordered} angle for $j$-th user]{\includegraphics[width=0.25\textwidth]{./img/Angle_Distirbutions}
% \label{fig:angle_dist}}
% \caption{PDF of \textit{ordered} distance and angle of $j$-th user for the limited feedback based ordering strategies of Section~\ref{sec:feedback_noma} with $\Delta\,{=}\,5^{\circ}$ and $j\,{=}\,20$.}
% \label{fig:angle_distance_dist}
% \end{figure}
%In the next section, we derive analytical expressions for the outage sum rate of NOMA transmission considering different user ordering strategies.
% In particular, we consider only $i$-th and $j$-th users in our investigations, though results can be generalized to more than 2 NOMA users as well.
\section{Analysis of NOMA Outage Sum Rates} \label{sec:Analytical_Rate_derivation}
% Assuming that each user has its own QoS based target rate, we evaluate respective sum rates to investigate conditions to serve each user at least at its target rate.

In this section, we analytically investigate the outage sum rates of the NOMA transmission described in Section~\ref{sec:Outage_Prob_Sum_Rates} under the limited feedback based ordering strategies of Section~\ref{sec:feedback_noma} and the full CSI based ordering. Without loss of generality, we consider two NOMA users (i.e., $K_{\rm N}\,{=}\,2$) in the analysis though the results can be generalized to multiple NOMA users, as well. In particular, in \cite{3GPP16NOMA_LTE} 3GPP has studied NOMA implementation for
LTE Release-13 under the name Multi-User Superposed Transmission (MUST). As defined in that, MUST-near UE (strong user) needs to be informed about paired MUST-far UE (weak user) in \emph{physical downlink shared channel} (PDSCH) along with their transmission power allocations whereas MUST-far UE needs to know only its own power allocation.

\subsection{Preliminaries for Outage Sum Rate Analysis}\label{sec:prelim_outage_analysis}
We assume that NOMA transmission targets $i$-th and $j$-th users with $i\,{>}\,j$, which are designated as the weaker and stronger users, respectively. In order to start transmission, there should be at least $j$ users (i.e., $K\,{\geq}\,j$), while NOMA transmission requires $K\,{\geq}\,i$ to have both $i$-th and $j$-th users simultaneously. Whenever we have $j\,{\leq}\,K\,{<}\,i$, there is only $j$-th user available and all resources are allocated to this user as per single user transmission approach.

Having defined the notations $\mathcal{S}_{K_{1}}$ and $\mathcal{S}_{K_{2}}$ in Section~\ref{sec:Outage_Prob_Sum_Rates}, we further define $\mathcal{S}_{K_{3}}:\left\lbrace K  \,|\, K \,{\geq}\, i \right\rbrace$ here, as well, to represent the set of $K$ values specifically enabling NOMA transmission. Note that sum rate calculation for $\mathcal{S}_{K_{1}}$ is performed using \eqref{eq:sum_rate_NOMA_singleK} whereas that for $\mathcal{S}_{K_{2}}$ and $\mathcal{S}_{K_{3}}$ is achieved employing \eqref{eq:sum_rate_NOMA_setK}, where both formulations make use of the conditional outage probability in \eqref{eq:Outage_k_th_user_2}. Unless otherwise stated, $\bar{\theta}$ is set to $0$ in the rest of the paper, which corresponds to the choice of a proper coordinate system as discussed in Section~\ref{sec:Outage_Prob_Sum_Rates}.

% In addition, the sum rate corresponding to $K\,{\geq}\,j$ can be obtained by summing up individual rate terms of $\mathcal{S}_{K_{2}}$ and $\mathcal{S}_{K_{3}}$.

Following the approach in~\cite{NadisankaTCoM}, the conditional outage probability of the $k$-th user given in \eqref{eq:Outage_k_th_user_2} can be expressed as
\begin{align} \label{eq:outage_probability_general_exp} \footnotesize
{\rm P}_{k|\mathcal{S}_{K}}^{\rm o} = \int\limits_{u_{\rm min}}^{u_{\rm max}} \int\limits_{L_1}^{L_2} &{\rm Pr} \left\lbrace \left| \textbf{h}_k^{\rm H}\textbf{b} \right|^2  < \eta_k \,{\big|}\, d_k\,{=}\,r,\,g({\theta_k})\,{=}\,u \right\rbrace
\nonumber \\
&\hfill \times f_{d_k|\mathcal{S}_{K}}(r) f_{g(\theta_k)|\mathcal{S}_{K}}(u) \dd r \dd u ,
\end{align}
where $u_{\rm min}\,{=}\,\min (g(\theta_k))$, $u_{\rm max}\,{=}\,\max (g(\theta_k))$, and $\eta_k$ is the variable to be specified for each NOMA user in the following. In \eqref{eq:outage_probability_general_exp}, $f_{d_k|\mathcal{S}_{K}}(r)$ represents the distance distribution of the $k$-th user, and $f_{g(\theta_k)|\mathcal{S}_{K}}(u)$ captures distribution of either angle or Fej\'er Kernel (as a function of angle) of the $k$-th user. Note that these two PDFs are written separately in \eqref{eq:outage_probability_general_exp} as per Lemma~\ref{the:unordered_ang_dist}, and will be derived in the subsequent sections for each NOMA user and ordering strategy separately.

When the outage probabilities are obtained individually for particular $K$ values as represented by $\mathcal{S}_{K_{1}}$ (as in the full CSI or Fej\'er Kernal based orderings), the outage sum rate can be computed directly using \eqref{eq:outage_probability_general_exp} in \eqref{eq:sum_rate_NOMA_singleK}. On the other hand, if the outage probabilities are computed for a range of $K$ values jointly as for $\mathcal{S}_{K_{2}}$ and $\mathcal{S}_{K_{3}}$ (as in the distance or angle based orderings), we need to first determine unconditional outage probabilities ${\rm P}_{k}^{\rm o}$ using \eqref{eq:outage_probability_general_exp} for each NOMA user $k$, and then compute \eqref{eq:sum_rate_NOMA_setK} accordingly.

Note that $i$-th user is present only for $\mathcal{S}_{K_3}$ where $K\,{\geq}\,i$, and the respective unconditional outage probability is therefore given considering \eqref{eq:sum_rate_NOMA_setK} as follows
\begin{align} \label{eq:Outage_Sk3_i_th_user}
 {\rm P}_{i}^{\rm o}{=}1{-} {\rm Pr} \{\mathcal{S}_{K_3}\}\big( 1 {-} {\rm P}_{i|\mathcal{S}_{K_3}}^{\rm o} \big).
\end{align}
Since $j$-th user is present for both $\mathcal{S}_{K_2}$ and $\mathcal{S}_{K_3}$, the desired unconditional outage probability is obtained similarly as, \vspace{-1em}

\small
\begin{align}\label{eq:Outage_Sk2_3_j_th_user}
 {\rm P}_{j}^{\rm o}{=} 1{-}\left[{\rm Pr} \left\lbrace \mathcal{S}_{K_2} \right\rbrace  \big( 1 {-} {\rm P}_{j|\mathcal{S}_{K_2}}^{\rm o} \big) {+} \; {\rm Pr} \{\mathcal{S}_{K_3}\} \big( 1 {-} {\rm P}_{j|\mathcal{S}_{K_3}}^{\rm o} \big) \right].
\end{align} \normalsize
Note that the variable $\eta_k$ in \eqref{eq:outage_probability_general_exp} is given for the $i$-th user as $\eta_i\,{=}\,\frac{\epsilon_i}{P_{\rm Tx}/N_0}$, whereas that for $j$-th user is either $\eta_j\,{=}\,\frac{\epsilon_j}{P_{\rm Tx}/N_0}$ or $\eta_j\,{=}\,\max\left\lbrace \frac{\epsilon_i/(P_{\rm Tx}/N_0)}{\beta_i^2{-}\beta_j^2\epsilon_i} ,\, \frac{\epsilon_j}{(P_{\rm Tx}/N_0)\beta_j^2} \right\rbrace$ depending on $\mathcal{S}_{K_2}$ or $\mathcal{S}_{K_3}$, respectively~\cite{NadisankaTCoM}. Note also that, since $K\,{\geq}\,j$ is initially assumed for a valid transmission, we need to normalize the outage sum rates by ${\rm Pr}\{K\,{\geq}\,j \}$ to obtain final rate results.

\subsection{Outage Probability for Fej\'er Kernel Based Ordering}
We first consider Fej\'er Kernel ordering in \eqref{eq:Fejer_ordering} to derive $f_{d_k|\mathcal{S}_{K}}(r)$ and $f_{g(\theta_k)|\mathcal{S}_{K}}(u)$ for each NOMA user, and then compute \eqref{eq:outage_probability_general_exp}. As explained in Section~\ref{sec:effect_ordering}, since Fej\'er Kernel is a function of angle, the ordered distance follows its unordered distribution while the distribution of Fej\'er Kernel (and equivalently that of angle) alters accordingly after ordering. Hence, we first give the distribution of the unordered distance, and then define the ordered distribution of Fej\'er Kernel in the following theorems.

\begin{lemma}\label{the:Unordered_distance_pdf}
Given the user region in Section~\ref{sec:Sys_Model}, the PDF of the unordered user distance is \begin{align} \label{eq:PDF_unordered_distance}
f_d(r) = \frac{ 2 r}{(L_2^2\,{-}\,L_1^2)}.
\end{align}
\end{lemma}
\begin{IEEEproof}
See Appendix~\ref{app:PDF_Unordered_user_dist}.
\end{IEEEproof}

\begin{figure}[!t]
\vspace{-2em}
\centering
\includegraphics[width=0.45\textwidth]{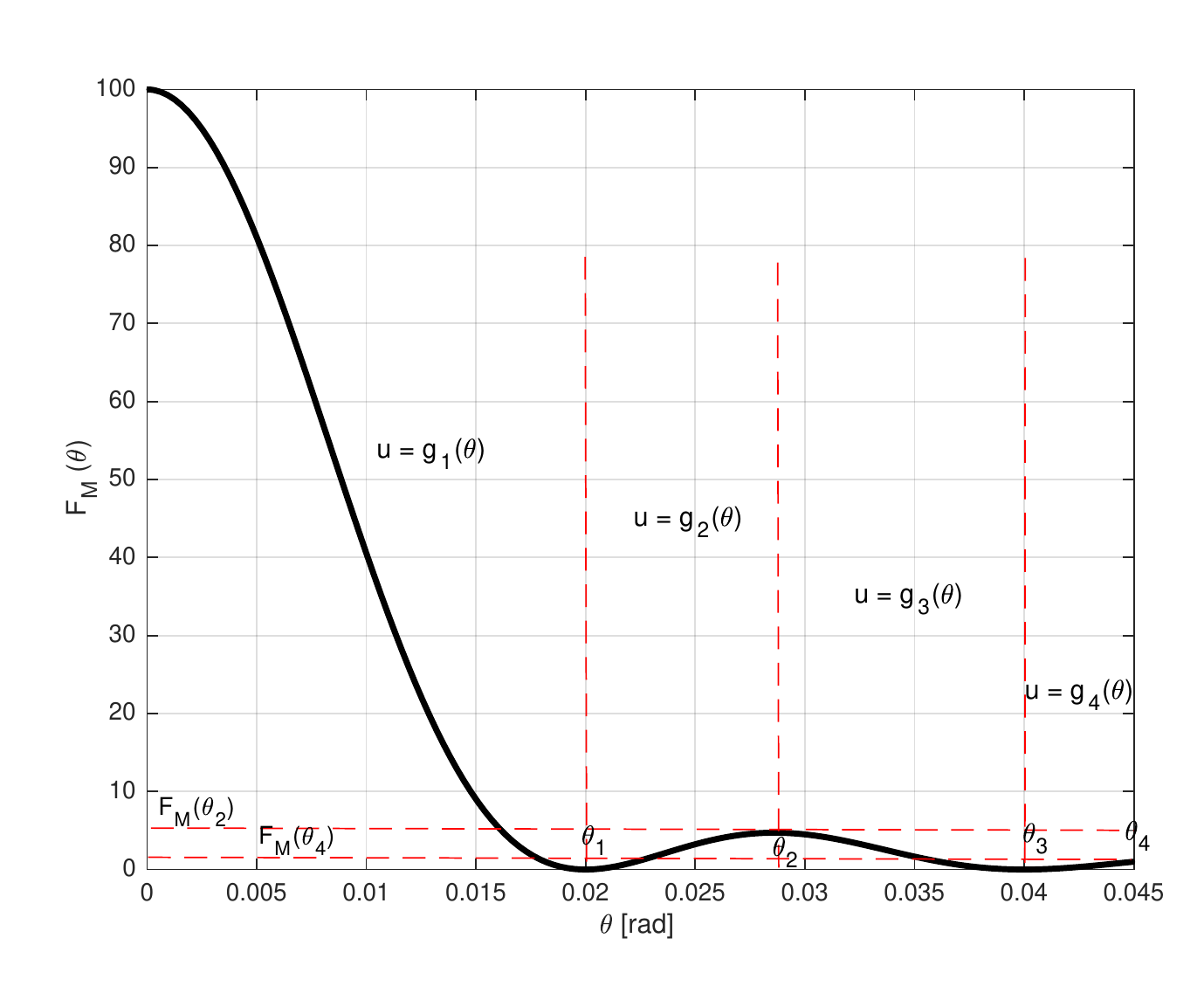}\vspace{-1mm}
\caption{Fej\'er Kernel function (taking values $u$) divided into four regions having a monotonic variation in $\theta$ with $\Delta\,{=}\,5^{\circ}$. Respective functions over each region are $g_1(\theta)$ for $0 \,{\leq}\, \theta \,{\leq}\, \theta_1$, $g_2(\theta)$ for $\theta_1 \,{\leq}\, \theta \,{\leq}\, \theta_2$, $g_3(\theta)$ for $\theta_2 \,{\leq}\, \theta \,{\leq}\, \theta_3$ and $g_4(\theta)$ for $\theta_3 \,{\leq}\, \theta \,{\leq}\, \theta_4$.}
\label{fig:Fejer-Kernel_func}
\end{figure}

\begin{theorem} \label{the:Ordered_Fejer_Kernel_pdf}
The PDF of the ordered Fej\'er Kernel denoted for the $k$-th user as $U_k$ is given by
\begin{align} \label{eq:PDF_ordered_Fejer_Kernel}
f_{U_k}(z) = c_k \frac{{\rm d}F_U(z)}{{\rm d}z}\left(F_U(z)\right)^{k-1}\left(1-F_U(z)\right)^{(K-k)},
\end{align}
where $U$ stands for the unordered Fej\'er Kernel and $c_k \,{=}\, \frac{K!}{(k-1)!(K-k)!}$. In addition, the CDF of unordered Fej\'er Kernel in \eqref{eq:PDF_ordered_Fejer_Kernel} can be given specifically for $M\,{=}\,100$ and $\Delta\,{=}\,5^{\circ}$ as follows \vspace{-1em}

\small
\begin{align} \label{eq:CDF_unordered_Fejer_Kernel}
\vspace{-2em}
&F_U(u)=
\nonumber \\
&\begin{cases}
\displaystyle 1, \hfill  u\geq 100, \\
\displaystyle 1{-}\frac{g_1^{-1}(u)}{\Delta/2}, \hfill {\rm F}_M(\theta_2) \leq u \leq 100, \\
\displaystyle \frac{g_2^{-1}(u) {-} g_1^{-1}(u){-}g_3^{-1}(u){+}\theta_4}{\Delta/2}, {\rm F}_M(\theta_4) \leq u \leq {\rm F}_M(\theta_2), \\
\displaystyle \frac{g_2^{-1}(u) {-} g_1^{-1}(u){-}g_3^{-1}(u){+}g_4^{-1}(u)}{\Delta/2}, \hfill u \leq {\rm F}_M(\theta_4),
\end{cases}
\end{align} \normalsize
where the functions $\{g_l(u)\}_{l{=}1}^{4}$ are obtained by partitioning Fej\'er Kernel into smaller parts changing monotonically with respect to $\theta$ as shown in Fig.~\ref{fig:Fejer-Kernel_func}, and $\{\theta_l\}_{l{=}1}^{4}$ are roots of $\partial {\rm F}_M(\theta) / \partial \theta$.
\end{theorem}

\begin{IEEEproof}
See Appendix~\ref{app:CDF_Unordered_Fejer_Kernel}.
\end{IEEEproof}

Note that the approach of Theorem~\ref{the:Ordered_Fejer_Kernel_pdf} in deriving \eqref{eq:CDF_unordered_Fejer_Kernel} can be generalized to any choice of $M$ and $\Delta$. For example, we obtain $F_U(u)\,{=}\,1\,{-}\,\frac{g_1^{{-}1}(u)}{\Delta/2}$ for $\Delta\,{=}\,1^{\circ}$ keeping $M$ the same. Finally, substituting $f_{d_k|\mathcal{S}_{K}}(r)$ and $f_{g(\theta_k)|\mathcal{S}_{K}}(u)$ with $f_d(r)$ and $f_{U_k}(z)$, respectively, and employing the distribution of $\left| \textbf{h}_k^{\rm H}\textbf{b} \right|^2$ (exponential)~\cite{NadisankaTCoM}, the outage in \eqref{eq:outage_probability_general_exp} becomes

\footnotesize
\begin{align}
&{\rm P}_{k|\mathcal{S}_{K_1}}^{\rm o} {=} \int\limits_{u_{\rm min}}^{u_{\rm max}} \int\limits_{L_1}^{L_2}  \Bigg( 1 {-} e^{\Big( \frac{-\eta_k{\textrm{PL} \big(\sqrt{r^2 {+} h^2}\big)}}{u} \Big)}  \Bigg) f_{U_k}(u)f_d(r) \dd r \dd u, \label{eq:Outage_Fejer_Kernel_1}\\
&= \int\limits_{0}^{1} \int\limits_{L_1}^{L_2}  \Bigg(1 {-} e^{ \Big( \frac{-\eta_k{\textrm{PL} \big(\sqrt{r^2 {+} h^2}\big)}}{F_U^{-1}(q)} \Big)}  \Bigg) c_k q^{k{-}1} (1{-}q)^{K{-}k}  f_d(r) \dd r \dd q, \label{eq:Outage_Fejer_Kernel}
\end{align} \normalsize
where $k\,{\in}\,\{i,j\}$. Note that, \eqref{eq:Outage_Fejer_Kernel} is obtained by substituting $q=F_U(u)$ in \eqref{eq:PDF_ordered_Fejer_Kernel} and \eqref{eq:Outage_Fejer_Kernel_1}, and adjusting outer integration limit accordingly. Note also that \eqref{eq:Outage_Fejer_Kernel} does not need the PDF of the ordered Fej\'er Kernel distribution, but the inverse of the unordered CDF only, which is easier to compute.

\subsection{Outage Probability for Angle Based Ordering}
We now consider angle based ordering given in \eqref{eq:angle_ordering}, and derive the PDFs $f_{d_k|\mathcal{S}_{K}}(r)$ and $f_{g(\theta_k)|\mathcal{S}_{K}}(u)$ for each NOMA user to compute the analytical outage probability in \eqref{eq:outage_probability_general_exp}. Similar to the Fej\'er Kernel ordering, the ordered distance follows its unordered distribution in this ordering strategy. However, the distribution of angle (and hence Fej\'er Kernel) alters since the ordering is performed with respect to angle, as discussed in Section~\ref{sec:effect_ordering}. Having derived the unordered distance distribution earlier in Lemma~\ref{the:Unordered_distance_pdf}, we derive the PDF of the $k$-th user angle, $\theta_k$, in the following theorems.

\begin{theorem}\label{lemma:Abs_Angle_pdf}
Assuming that the number of users $K$ takes values from $\mathcal{S}_{K_2}$ such that $j \,{\leq}\, K \,{<}\, i$, the PDF of the ordered \textit{absolute} angle $\tilde{\theta}_k \,{=}\,|\theta_k|$ for the $k$-th user is given as \vspace{-1em}

\small
\begin{align} \label{eq:PDF_j_less_K_less_i}
&f_{\tilde{\theta}_k|\mathcal{S}_{K_2}}(\theta) {=} \frac{ L}{\mathcal{C}}e^{-\frac{\Delta}{2} L}\frac{\left[\theta L \right]^{(k{-}1)}}{(k{-}1)!} \left\lbrace \sum \limits_{l{=}0}^{i{-}k{-}1}\frac{\left[(\frac{\Delta}{2}{-}\theta)L \right]^{l}}{l!} \right\rbrace; \theta{\geq} 0 ,
\end{align} \normalsize
where $\mathcal{C}\,{=}\,\sum\limits_{l=j}^{i{-}1} \frac{e^{{-}\frac{\Delta}{2} L} \left[\frac{\Delta}{2} L \right]^l}{l!}$ and $L\,{=}\,(L_2^2 - L_1^2)\lambda$. Furthermore, if $K\,{\in}\,\mathcal{S}_{K_3}$ such that $K \,{\geq}\, i$, we have \small
\begin{align} \label{eq:PDF_K_greater_j}
f_{\tilde{\theta}_k|\mathcal{S}_{K_3}}(\theta) &= \frac{ L}{\mathcal{C}}\frac{\left[\theta L \right]^{(k{-}1)}}{(k{-}1)!} \left\lbrace e^{-\theta L}{-} e^{-\frac{\Delta}{2} L} \sum \limits_{l=0}^{i{-}k{-}1}\frac{\left[(\frac{\Delta}{2}{-}\theta) L \right]^{l}}{l!} \right\rbrace
\nonumber \\
&\hspace{17em}; \theta\geq 0 ,
\end{align} \normalsize
where $\mathcal{C}\,{=}\, \left\lbrace 1-\sum\limits_{l=0}^{i{-}1} \frac{e^{{-}\frac{\Delta}{2} L} \left[\frac{\Delta}{2} L \right]^l}{l!} \right\rbrace$ and $L\,{=}\,(L_2^2 - L_1^2)\lambda$.
\end{theorem}\vspace{1em}

\begin{IEEEproof}
See Appendix~\ref{app:PDF_Ordered_Abs_Angle}.
\end{IEEEproof}

\begin{theorem} \label{theorem:Ordered_angle_pdf}
Considering the set of $K$ values for a valid transmission such that $K\,{\geq}\,j$ or equivalently $K\,{\in}\,\mathcal{S}_{K_2} \,{\cup}\, \mathcal{S}_{K_3}$, the PDF of the ordered angle, $\theta_k$, for the $k$-th user is given as
\begin{align}
f_{\theta_k|\mathcal{S}_{K_m}}(\theta) = \frac{1}{2}f_{\tilde{\theta}_k|\mathcal{S}_{K_m}}(|\theta|),
\end{align}
where $m=\{2,3\}$, and $f_{\tilde{\theta}_k|\mathcal{S}_{K_m}}$ is given in \eqref{eq:PDF_j_less_K_less_i} and \eqref{eq:PDF_K_greater_j}.
\end{theorem}
\begin{IEEEproof}
See Appendix~\ref{app:PDF_Ordered_Angle}.
\end{IEEEproof}
Similar to \eqref{eq:Outage_Fejer_Kernel_1}, incorporating $f_d(r)$ and $f_{\theta_k|\mathcal{S}_{K_m}}(\theta)$ from Lemma~\ref{the:Unordered_distance_pdf} and Theorem~\ref{theorem:Ordered_angle_pdf} into \eqref{eq:outage_probability_general_exp}, the desired outage probability for NOMA users $i$ and $j$ is given as

\footnotesize
\begin{align} \label{eq:Outage_Abs_Angle}
{\rm P}_{k|\mathcal{S}_{K_m}}^{\rm o}{=}\int\limits_{{-}\frac{\Delta}{2}}^{\frac{\Delta}{2}} \int\limits_{L_1}^{L_2}  \Bigg( 1 {-} e^{ \Big( \frac{-\eta_k{\textrm{PL} \big(\sqrt{r^2 {+} h^2}\big)}}{ {{\rm F}_M(\theta)}} \Big) }  \Bigg) f_{\theta_k|\mathcal{S}_{K_m}}(\theta) f_d(r) \dd r \dd \theta ,
\end{align} \normalsize where $k\,{\in}\,\{i,j\}$ and $m\,{=}\,\{2,3\}$.

\subsection{Outage Probability for Distance Based Ordering}
We now consider distance ordering given in \eqref{eq:distance_ordering}, and derive $f_{d_k|\mathcal{S}_{K}}(r)$ and $f_{g(\theta_k)|\mathcal{S}_{K}}(u)$ to compute \eqref{eq:outage_probability_general_exp} accordingly. When we order users based on the distance information, the distribution of angle (or equivalently Fej\'er Kernel) does not change while distance distribution alters accordingly, as discussed in Section~\ref{sec:effect_ordering}. The unordered angle actually follows a uniform distribution, and the PDF of the ordered distance is given in the following.

\begin{lemma}
Assuming that the number of users $K$ takes values from $\mathcal{S}_{K_2}$ such that $j \,{\leq}\, K \,{<}\, i$, the PDF of the ordered distance $d_k$ for the $k$-th user is given as
\begin{align} \label{eq:ordered_distance_pdf_sk2}
f_{d_k|\mathcal{S}_{K_2}}(r) = \frac{ \Delta \lambda r}{\mathcal{C}}&e^{-\frac{\Delta}{2}L}\frac{\left[\frac{\Delta}{2}(r^2 {-} L_1^2)\lambda \right]^{(k-1)}}{(k-1)!}
\nonumber \\
& \hspace{2em} \times \left( \sum \limits_{l=0}^{i-k-1}\frac{\left[\frac{\Delta}{2}(L_2^2 {-} r^2)\lambda \right]^{l}}{l!} \right)
\end{align}
where $\mathcal{C}\,{=}\,\sum\limits_{l=j}^{i{-}1} \frac{e^{{-}\frac{\Delta}{2}L} \left[\frac{\Delta}{2}L \right]^l}{l!}$ and $L\,{=}\,(L_2^2 {-} L_1^2)\lambda$.
\end{lemma}

\begin{IEEEproof}
See~\cite{NadisankaTCoM}.
\end{IEEEproof}

\begin{theorem}
Further, when the number of users $K$ obeys $\mathcal{S}_{K_3}$ such that $K \,{\geq}\, i$, the PDF of the ordered distance $d_k$ for the $k$-th user is given as
\begin{align} \label{eq:ordered_distance_pdf_sk3}
&f_{d_k|\mathcal{S}_{K_3}}(r)= \frac{ \Delta \lambda r}{\mathcal{C}}\frac{\left[\frac{\Delta}{2}(r^2 - L_1^2)\lambda \right]^{(k-1)}}{(k-1)!}
\nonumber \\
&\hspace{1em} \times \left(e^{-\frac{\Delta}{2}(r^2 - L_1^2)\lambda} -  e^{-\frac{\Delta}{2}L}\sum \limits_{l=0}^{i-k-1}\frac{\left[\frac{\Delta}{2}(L_2^2 - r^2)\lambda \right]^{l}}{l!} \right)
\end{align}
where $\mathcal{C}\,{=}\,1-\sum\limits_{l=0}^{i{-}1} \frac{e^{{-}\frac{\Delta}{2}L} \left[\frac{\Delta}{2}L \right]^l}{l!}$ and $L\,{=}\,(L_2^2 - L_1^2)\lambda$.
\end{theorem}

\begin{IEEEproof}
See Appendix~\ref{app:PDF_Ordered_Distance_K_greater_i}.
\end{IEEEproof}

Employing $f_{d_k|\mathcal{S}_{K_m}}(r)$ given in \eqref{eq:ordered_distance_pdf_sk2} and \eqref{eq:ordered_distance_pdf_sk3}, and $f_{g(\theta_k)|\mathcal{S}_{K}}(u)$ being equal to $\frac{1}{\Delta}$ within the interval $\left[{-}\frac{\Delta}{2},\frac{\Delta}{2}\right]$, the desired outage probability in \eqref{eq:outage_probability_general_exp} can be given as \vspace{-1em}

\small
\begin{align} \label{eq:Outage_Distance}
{\rm P}_{k|\mathcal{S}_{K_m}}^{\rm o} {=} \frac{1}{\Delta} \int\limits_{{-}\frac{\Delta}{2}}^{\frac{\Delta}{2}} \int\limits_{L_1}^{L_2}  \Bigg( 1 {-} e^{\Big( \frac{-\eta_k{\textrm{PL} \big(\sqrt{r^2 {+} h^2}\big)}}{ {{\rm F}_M(\theta)}} \Big)}  \Bigg) f_{d_k|\mathcal{S}_{K_m}}(r) \dd r \dd \theta ,
\end{align} \normalsize where $k\in \{i,j\}$ and $m=\{ 2, 3 \}$.

\subsection{Outage Probability for Full CSI Based Ordering}
In this section, we provide outage probabilities for full CSI (or, equivalenttly effective channel gain) based ordering to provide a comparison with the performance of limited feedback schemes considered so far. Note that the CDF of the \textit{unordered} effective channel gain for the scenario in Section~\ref{sec:Sys_Model} is given as~\cite{NadisankaTCoM} \vspace{-1em}

\small
\begin{align}\label{eq:cdf_unordered_fullcsi}
F_{\pi}(y) {=} \int\limits_{{-}\frac{\Delta}{2}}^{\frac{\Delta}{2}} \int\limits_{L_1}^{L_2}  \Bigg( 1 {-} e^{ \Big( \frac{-y{\textrm{PL} \big(\sqrt{r^2 {+} h^2}\big)}}{ {{\rm F}_M(\theta)}} \Big)}  \Bigg) \frac{r}{\frac{\Delta}{2}(L_2^2 {-} L_1^2)} \dd r \dd \theta,
\end{align} \normalsize
and respective \textit{ordered} CDF can be computed using order statistics, which is actually equivalent to the conditional outage probability in \eqref{eq:outage_probability_general_exp}. Note that, \eqref{eq:cdf_unordered_fullcsi} makes use of the fact that users are homogeneously distributed within the user region with the area $\mathcal{A}\,{=}\,\frac{\Delta}{2}(L_2^2 {-} L_1^2)$, and respective PDF of the user location is therefore  $\frac{r}{\mathcal{A}}$ in polar coordinates.

\section{Numerical Results} \label{sec:Numerical_results}

In this section, we study in detail the achievable outage sum rates with NOMA and OMA transmissions for the scenario captured in Section~\ref{sec:Sys_Model}. In particular, by using the derived analytical expressions in Section~\ref{sec:Analytical_Rate_derivation} and through extensive Monte Carlo simulations we investigate the NOMA performance with various feedback schemes and user ordering strategies. Further, we study the impact of user region geometry on the NOMA feedback scheme. Unless stated otherwise, we assume that $L_{1}\,{=}\,85$~m, $L_{2}\,{=}\,100$~m, $M\,{=}\,100$, $\Delta\,{\in}\,\left\lbrace 1^{\circ},5^{\circ}\right\rbrace$, and $\bar{\theta}\,{=}\,0^{\circ}$. Users are deployed based on HPPP with $\lambda\,{=}\,1$ and the noise, $N_0{=}-35$~dBm. QoS based target rates for $j$-th and $i$-th users are, $\overline{R}_j\,{=}\,6$~bits per channel use (BPCU) and $\overline{R}_i\,{=}\,0.5$~BPCU \cite{NadisankaTCoM, Ding17PoorRandBeamforming}, respectively while the power allocation ratios for $j$-th and $i$-th users are set as $\beta_j^2\,{=}\,0.25$ and $\beta_i^2\,{=}\,0.75$. The PL model is assumed to be $\textrm{PL}(\sqrt{d_k^2 + h^2}) {=} 1 {+} \left(\sqrt{d_k^2 + h^2}\right)^{\gamma}$ with $\gamma{=}2$ \cite{Ding17PoorRandBeamforming, NadisankaTCoM}. Finally, the UAV-BS hovering altitude range is assumed to be $h\,{\in}\,[10,150]$~m, complying with regulations of authorities in charge \cite{FAARule}.

\subsection{Performance of Ordering Strategies: NOMA vs. OMA}

\begin{figure}[!t]
%\vspace{-2em}
\centering
\includegraphics[width=0.45\textwidth]{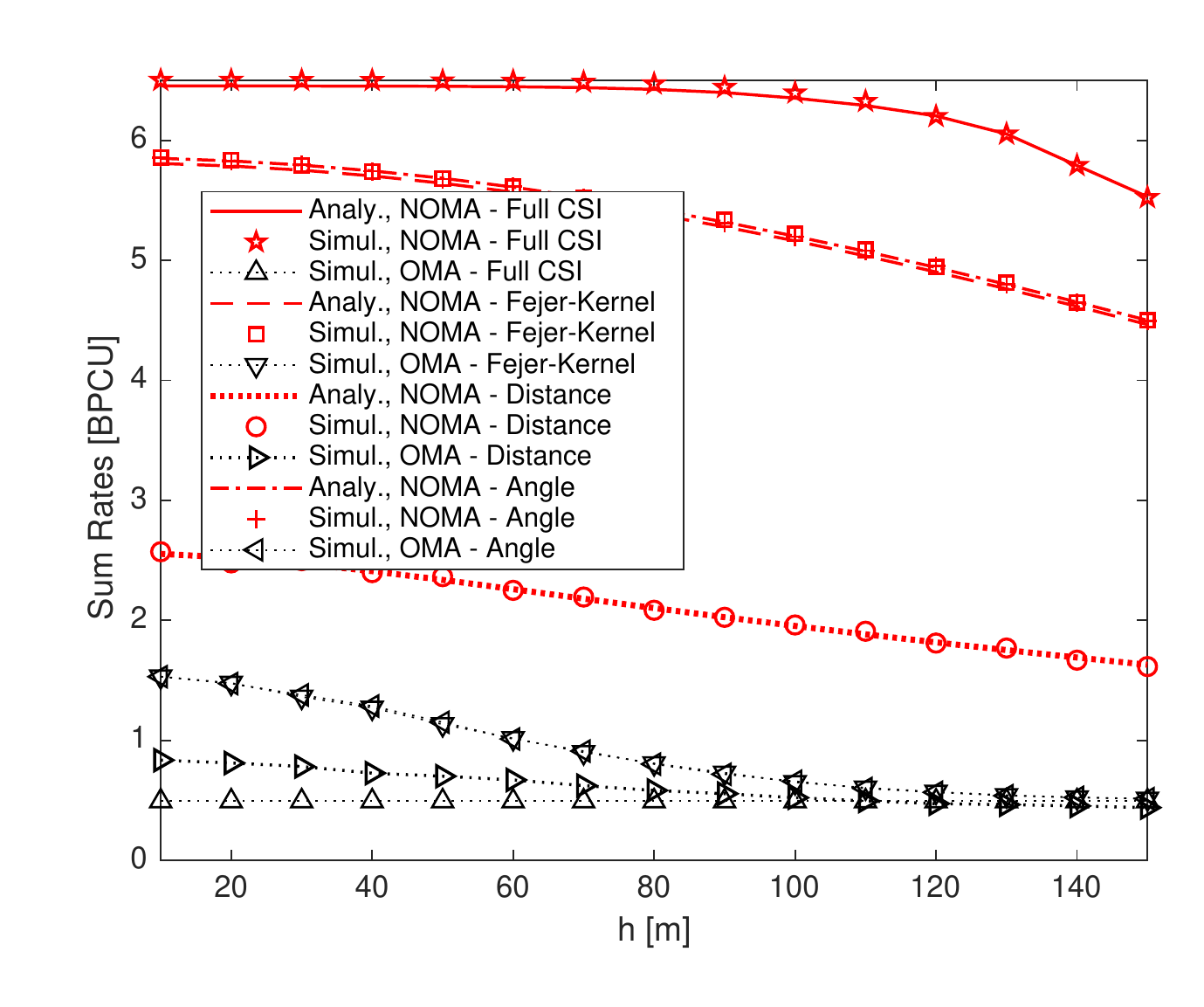}\vspace{0mm}
\caption{Sum rates for NOMA and OMA for full CSI, Fej\'er kernel, distance and angle based ordering strategies with $i\,{=}\,25$, $j\,{=}\,20$, $P_{\rm Tx}\,{=}\,20$~dBm, and $\Delta\,{=}\,5^{\circ}$.}
\label{fig:sumrate_distance_fejer_NOMA_OMA_j20_i25}
\end{figure}

\begin{figure}[!t]
%\vspace{-2em}
\centering
\subfloat[$i$-th user with $i\,{=}\,25$.]{\includegraphics[width=0.45\textwidth]{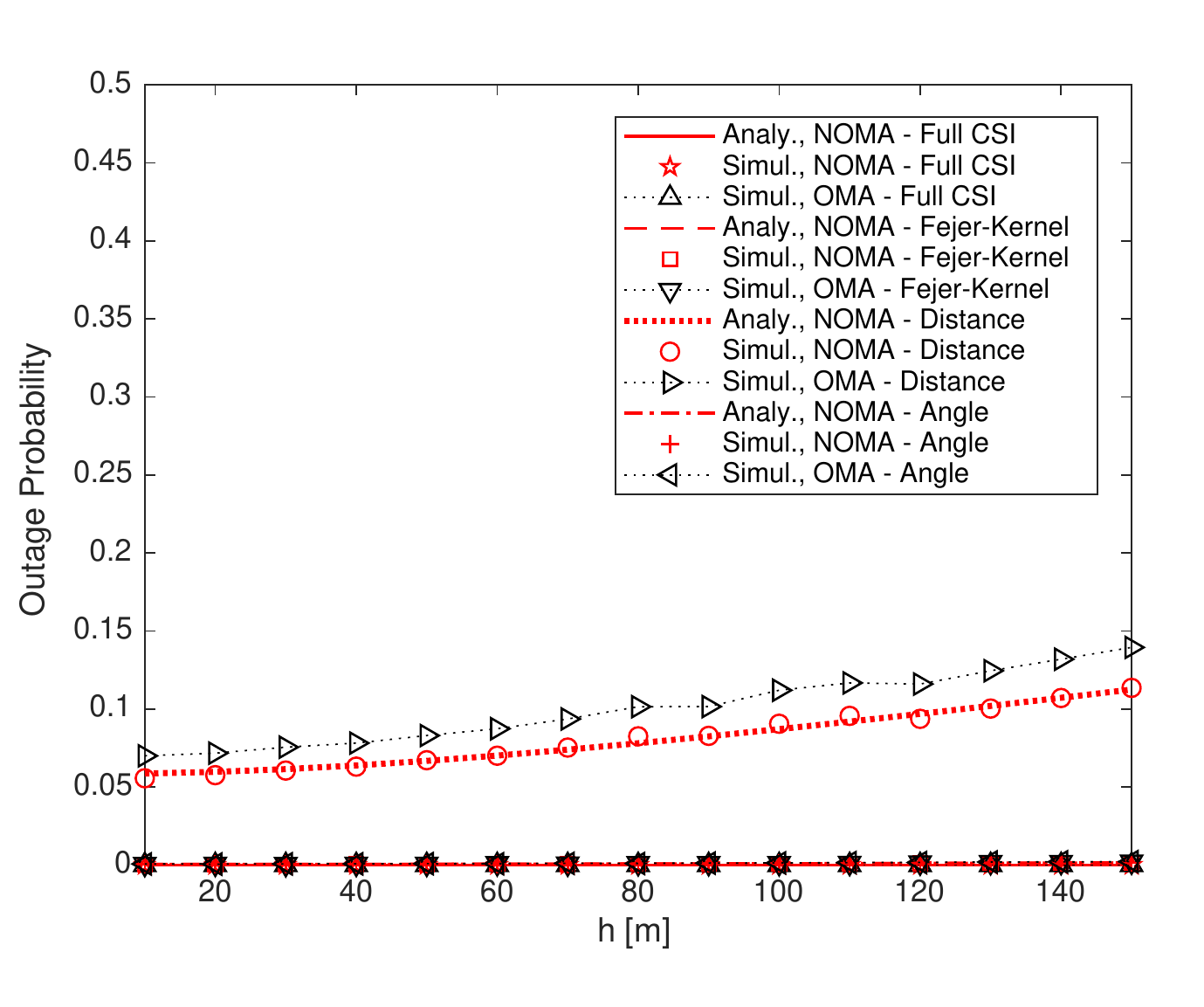}
\label{fig:outage_distance_j20_i25_ith}}\\
%\vspace{-1em}
\subfloat[$j$-th user with $j\,{=}\,20$.]{\includegraphics[width=0.45\textwidth]{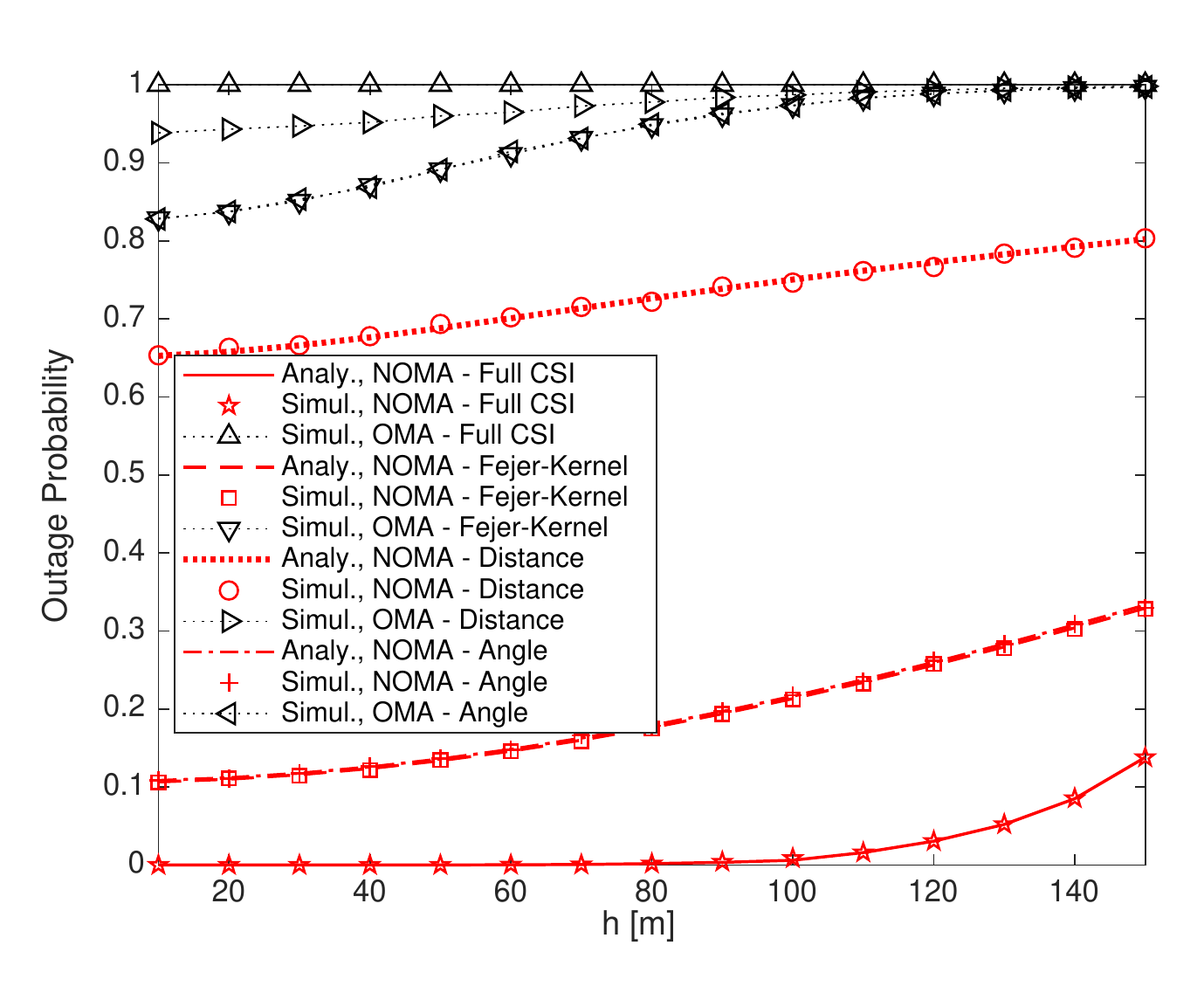}
\label{fig:outage_distance_j20_i25_jth}} \vspace{1em}
\caption{Outage probabilities for NOMA and OMA for full CSI, Fej\'er kernel, distance and angle based ordering strategies with $P_{\rm Tx}\,{=}\,20$~dBm and $\Delta\,{=}\,5^{\circ}$.}
\label{fig:outage_distance_j20_i25}
\end{figure}

In Fig.~\ref{fig:sumrate_distance_fejer_NOMA_OMA_j20_i25}, we present outage sum rates of OMA and NOMA along with varying UAV-BS altitudes for ordering criteria discussed in Section~\ref{sec:Analytical_Rate_derivation} considering $i\,{=}\,25$, $j\,{=}\,20$, $P_{\rm Tx}\,{=}\,20$~dBm, $\Delta\,{=}\,5^{\circ}$, and $P_{\rm Tx}\,{=}\,20$~dBm. The numerical results verify the derivations in Section~\ref{sec:Analytical_Rate_derivation} by showing a perfect match between analytical and simulation results. In addition, outage sum rate performance of NOMA outperforms that of OMA for all ordering criteria. We observe that sum rate performance of Fej\'er kernel and angle based ordering strategies are very similar (to be detailed in Section~\ref{sec:results_fejer_vs_angle}), and are significantly better than that of the distance based ordering. In addition, these results are also consistent with the outage probability results presented in Fig.~\ref{fig:outage_distance_j20_i25}, where outage probabilities for $j$-th user are much more apparent.

\begin{figure}[!t]
%\vspace{-2em}
\centering
\includegraphics[width=0.45\textwidth]{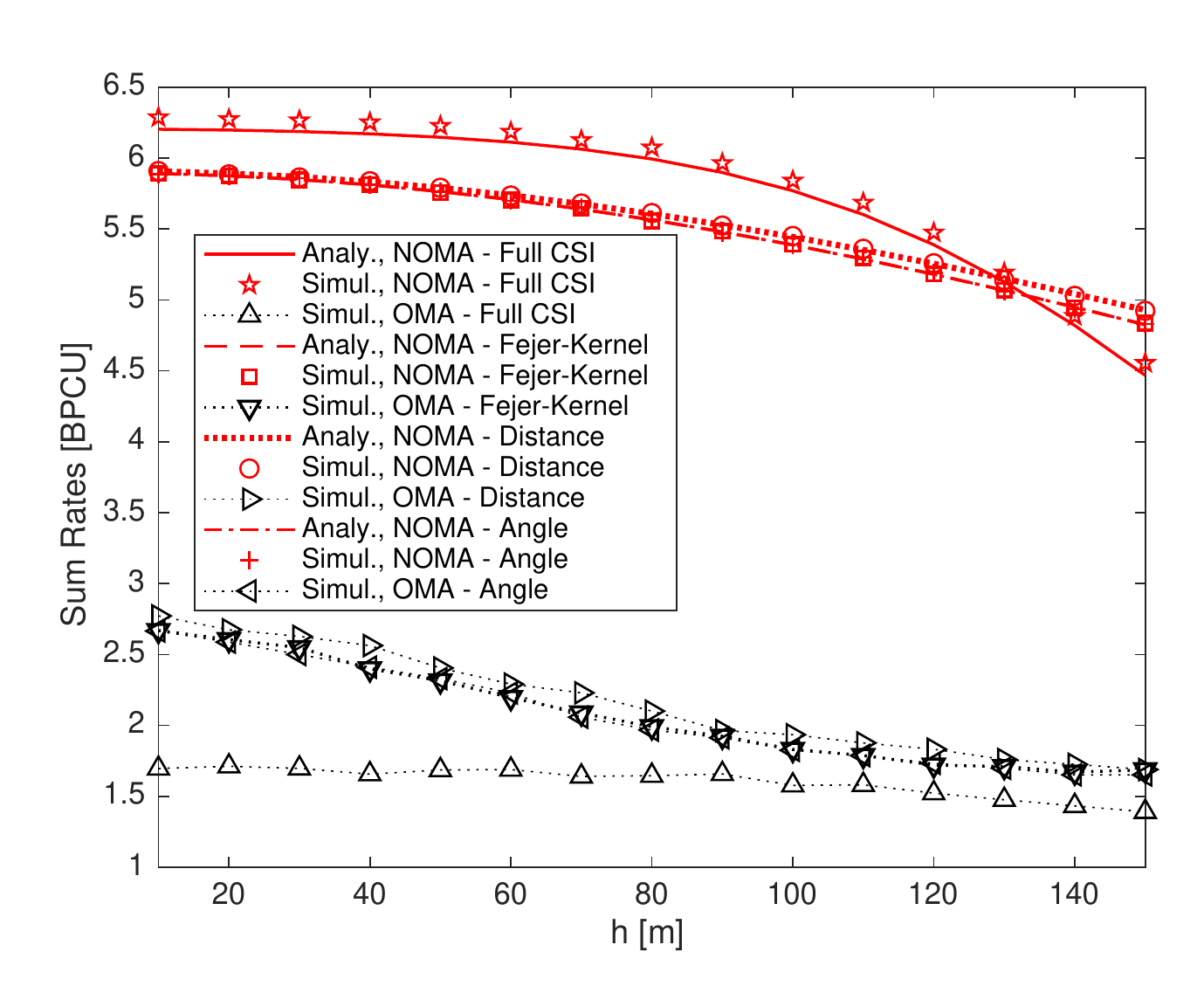}\vspace{0mm}
\caption{Sum rates for NOMA and OMA for full CSI, Fej\'er kernel, distance and angle based ordering strategies with $i\,{=}\,25$, $j\,{=}\,20$, $P_{\rm Tx}\,{=}\,20$~dBm, and $\Delta\,{=}\,1^{\circ}$.}
\label{fig:sumrate_distance_fejer_NOMA_OMA_j20_i25_1_deg}
\end{figure}

In Fig.~\ref{fig:sumrate_distance_fejer_NOMA_OMA_j20_i25_1_deg}, we generate the outage sum rates for the same setting of Fig.~\ref{fig:sumrate_distance_fejer_NOMA_OMA_j20_i25}, except for a narrower horizontal angle with $\Delta\,{=}\,1^{\circ}$. We observe that Fej\'er kernel and angle based ordering strategies lose their power for this specific setting, and the respective sum rate performances are very similar to that of the distance based ordering. The reason for this result lies in the fact that potential NOMA users become \textit{less distinctive} based on their angular information when $\Delta$ gets smaller. Hence, distance information might become a comparable or even more powerful feature in distinguishing different users within this setting. As a result, distance based ordering provides a relatively better power domain separation for NOMA transmission.

Note that in Fig.~6, we interestingly observe that the sum-rate performance of full CSI based ordering drops below that of the limited-feedback based ordering schemes after a $130\,\text{m}$ of flight altitude. As rigorously analyzed in Section~V.B of \cite{NadisankaTCoM}, the \textit{actual order} of any NOMA user with respect to full CSI (i.e., actual order $k$ means the user having the $k$th strongest channel) is very likely to be different than its order with respect to any limited-feedback information. Therefore, the actual orders of limited-feedback NOMA users are very likely not to be $i$ and $j$ any more, but rather take different values at each trial. Since the outage performance depends on the actual order of NOMA users, limited-feedback schemes can result in worse or better sum-rate performance depending on the specific setting. This is the reason lying under the sum-rate performance of full CSI based ordering being worse than that of the limited-feedback based ordering strategies after $130\,\text{m}$ UAV altitude.

\subsection{Fej\'er Kernel vs. Distance Based Ordering}
% \textcolor{red}{or should we tell angle vs. distance feedback?}}{\color{blue}Yavuz: Possible, but how to alter the title of the next section then?}
\label{sec:Perf_Rates_Dist_Fejer}

\begin{figure}[!t]
%\vspace{-1em}
\centering
\includegraphics[width=0.45\textwidth]{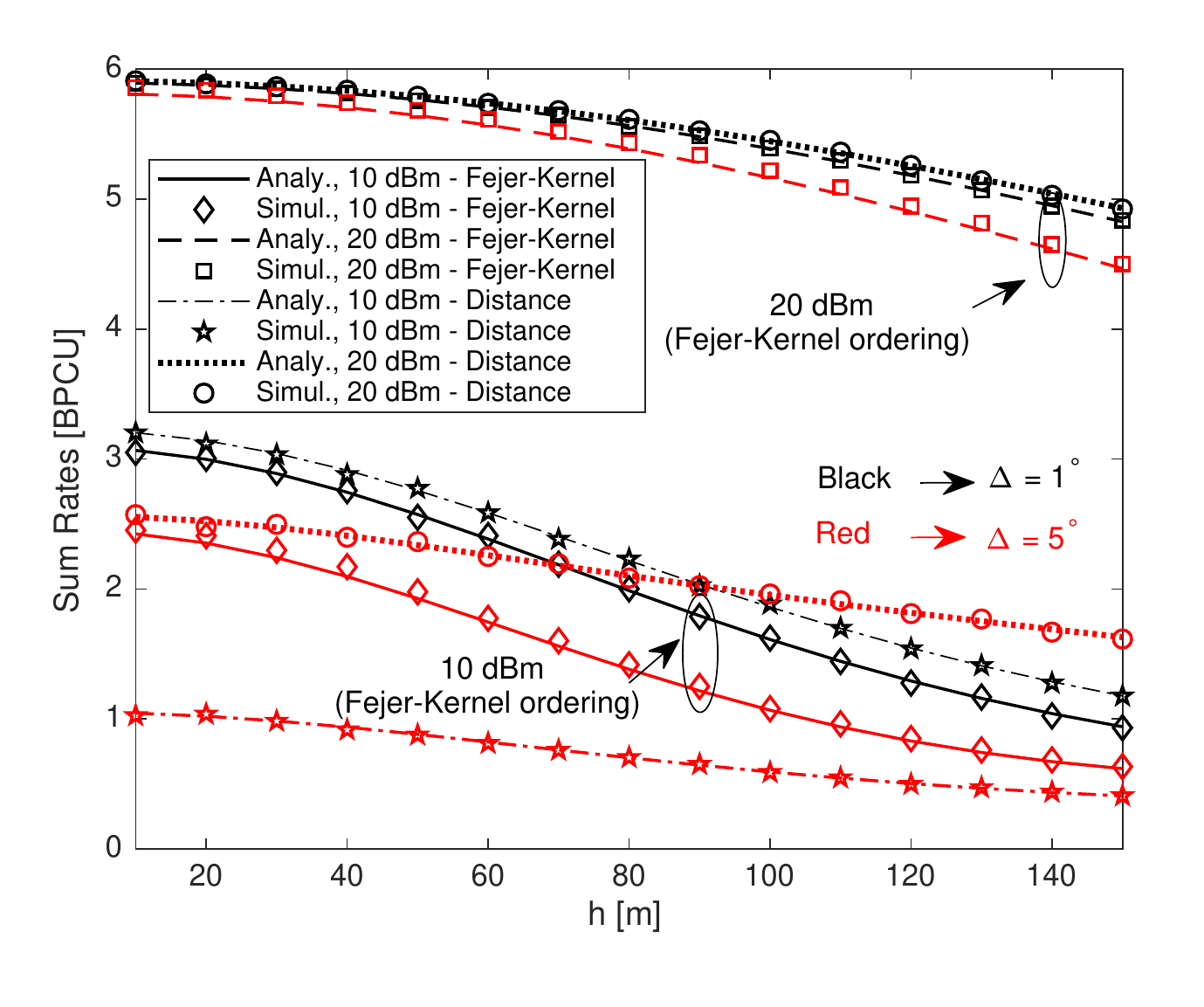}\vspace{-1mm}
\caption{NOMA sum rates for Fej\'er kernel and distance based ordering strategies with $i\,{=}\,25$, $j\,{=}\,20$, $P_{\rm Tx}\,{=}\,\{10,20\}$~dBm, and $\Delta\,{=}\,\{1^{\circ},5^{\circ}\}$.}
\label{fig:sumrate_distance_fejer_comparison_j20_i25}
\end{figure}

\begin{figure}[!t]
%\vspace{-1em}
\centering
\includegraphics[width=0.45\textwidth]{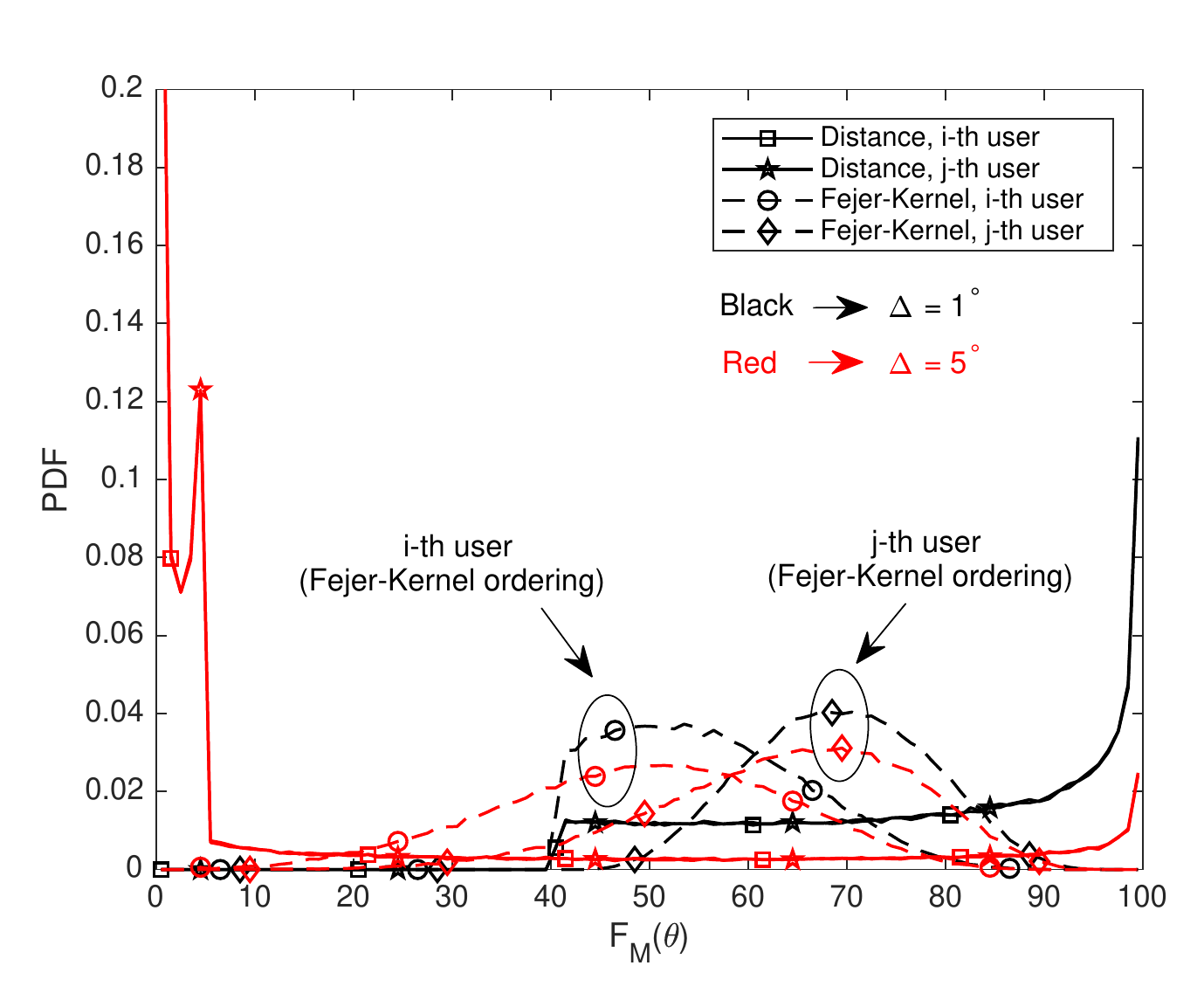}\vspace{-1mm}
\caption{PDF of Fej\'er kernel ${\rm F}_M(\theta)$ with the distance and Fej\'er kernel based ordering strategies for $i\,{=}\,25$, $j\,{=}\,20$, and $\Delta\,{=}\,\{1^{\circ},5^{\circ}\}$.}
\label{fig:PDF_Fejer_Kernel}
\end{figure}

In Fig.~\ref{fig:sumrate_distance_fejer_comparison_j20_i25} we capture NOMA outage sum rates for Fej\'er kernel and distance based user ordering strategies for $i\,{=}\,25$, $j\,{=}\,20$, $P_{\rm Tx}\,{=}\,\{10,20\}$~dBm, and $\Delta\,{=}\,\{1^{\circ},5^{\circ}\}$. From that, we can observe for both transmit power values fej\'er kernel based ordering outperforms distance based ordering for relatively larger $\Delta$ values. In particular, the sum rate performance of Fej\'er kernel based ordering at $P_{\rm Tx}\,{=}\,20$~dBm is significantly superior to that of the distance based ordering when $\Delta\,{=}\,5^{\circ}$ and this difference deteriorates dramatically as $\Delta$ gets smaller.

In order to give more insights on the NOMA sum rate behavior for the Fej\'er kernel and distance based ordering strategies, we depict the respective PDF of Fej\'er kernel, ${\rm F}_M(\theta)$ in Fig.~\ref{fig:PDF_Fejer_Kernel}. Note that the effective channel gain in \eqref{eq:Eff_channel_gain} is more sensitive to variations in ${\rm F}_M(\theta)$ than PL term, since ${\rm F}_M(\theta)$ values are much bigger in magnitudes than the PL values. We observe that ${\rm F}_M(\theta)\,{\in}\,[40,100]$ for both ordering strategies when $\Delta\,{=}\,1^{\circ}$, where the large ${\rm F}_M(\theta)$ values are more probable for the distance based ordering. As a result, distance based ordering tends to achieve slightly better outage sum rate performance for $\Delta\,{=}\,1^{\circ}$. In contrast, we have ${\rm F}_M(\theta)\,{\in}\,[10,90]$ for Fej\'er kernel based ordering when $\Delta\,{=}\,5^{\circ}$, while respective ${\rm F}_M(\theta)$ values for the distance based ordering are very likely to appear within $[0,7]$. As a result, Fej\'er kernel based ordering achieves superior rate performance compared to that of distance based ordering when $\Delta\,{=}\,5^{\circ}$.

% In Fig.~\ref{fig:sumrate_distance_fejer_comparison_j20_i25}, the effect of DL power budget on NOMA sum rates is investigated under distance and Fej\'er kernel based user ordering schemes with $i\,{=}\,25$, $j\,{=}\,20$, $\Delta\,{\in}\,\left\lbrace 1^{\circ},5^{\circ}\right\rbrace$, and $P_{\rm Tx}\,{=}\,\{10,20\}$~dBm. We observe the performance gap between distance and Fej\'er kernel based ordering schemes become larger for relatively small transmit power of $P_{\rm Tx}\,{=}\,10$~dBm, and that this gap is even more significant for the wide horizontal angle of $\Delta\,{=}\,5^{\circ}$.

\subsection{Fej\'er Kernel vs. Angle Based Ordering}\label{sec:results_fejer_vs_angle}

\begin{figure}[!t]
%\vspace{-1em}
\centering
\includegraphics[width=0.45\textwidth]{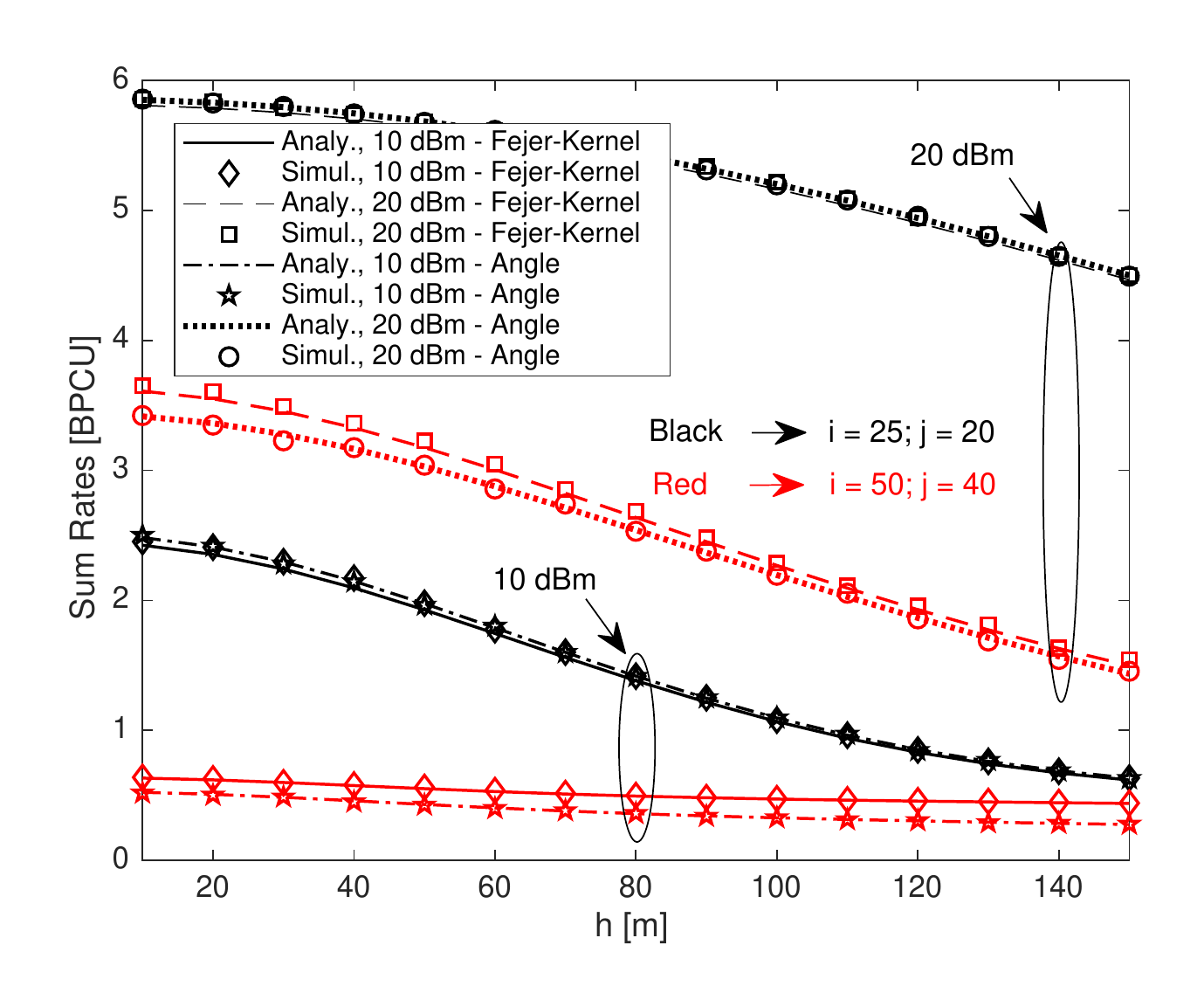}\vspace{-1mm}
\caption{NOMA sum rates for Fej\'er kernel and angle based ordering strategies with $\{i\,{=}\,25$, $j\,{=}\,20\}$, $\{i\,{=}\,50$, $j\,{=}\,40\}$, $P_{\rm Tx}\,{=}\,\{10,20\}$~dBm, and $\Delta\,{=}\,5^{\circ}$.}
\label{fig:sumrate_angle_fejer_comparison_j20_i25}
\end{figure}

\begin{figure}[!t]
%\vspace{-5mm}
\centering
\includegraphics[width=0.45\textwidth]{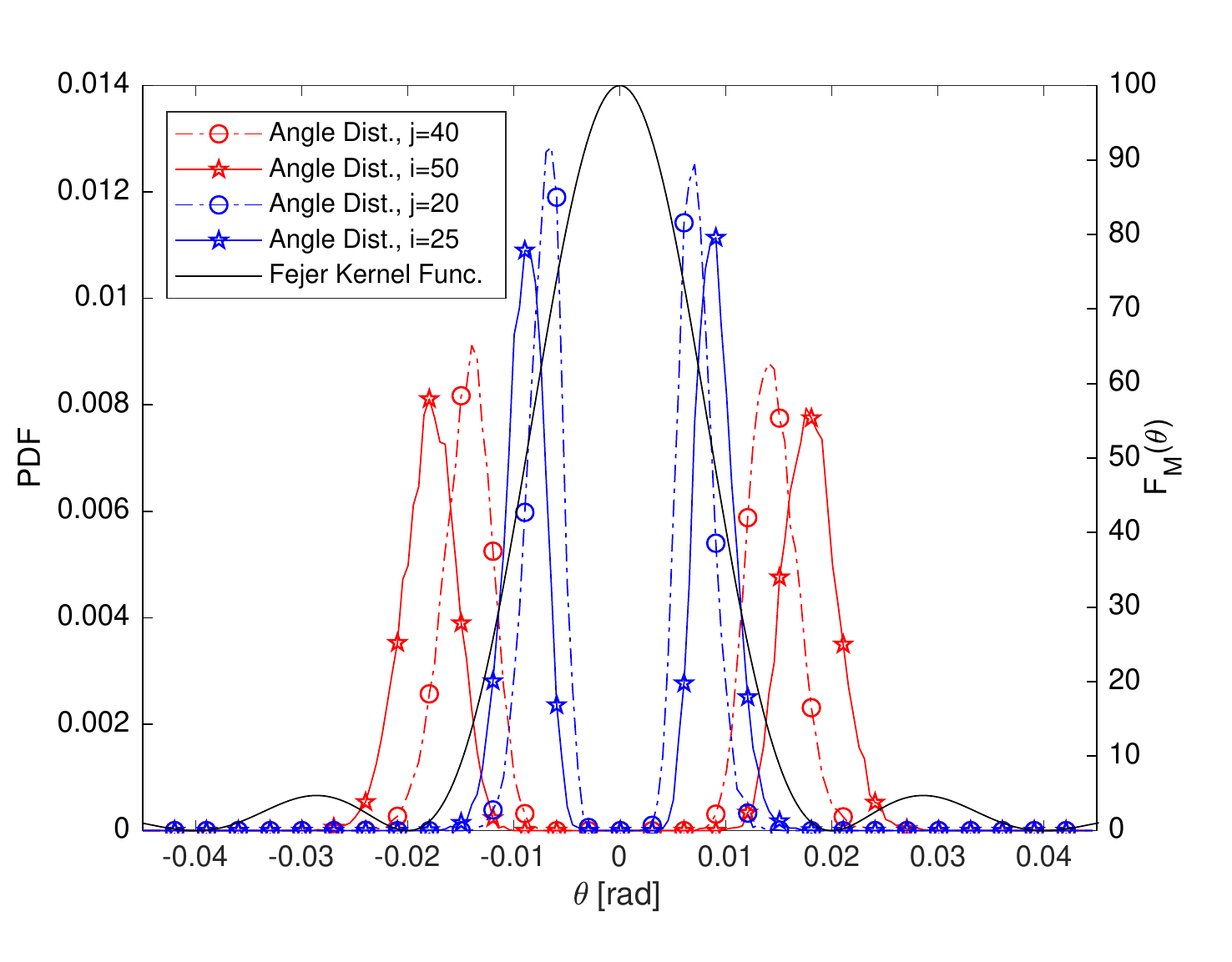}\vspace{-1mm}
\caption{PDFs of angle distribution for angle based ordering for $\Delta\,{=}\,5^{\circ}$ with the user pairs $\{i\,{=}\,25$, $j\,{=}\,20\}$ and $\{i\,{=}\,50$, $j\,{=}\,40\}$.}
\label{fig:PDF_Angle_Distribution}
\end{figure}

Achievable outage sum rates for NOMA considering Fej\'er kernel and angle based ordering are presented in Fig.~\ref{fig:sumrate_angle_fejer_comparison_j20_i25} along with varying altitudes for $\Delta\,{=}\,5^{\circ}$ and $P_{\rm Tx}\,{=}\,\left\lbrace 10,20 \right\rbrace$~dBm. In this case, we consider two different users pairs, 1) $i\,{=}\,25$, $j\,{=}\,20$, and 2) $i\,{=}\,50$, $j\,{=}\,40$. We observe that although both ordering strategies perform the same when $i\,{=}\,25$ and $j\,{=}\,20$, sum rate performance of the Fej\'er kernel based ordering becomes better as compared to angle based ordering when $i\,{=}\,50$, $j\,{=}\,40$. To investigate the reason behind this behavior, we plot the Fej\'er kernel function and the PDF of user angle $\theta_k$ in Fig.~\ref{fig:PDF_Angle_Distribution} considering angle based user ordering strategy.

We observe in Fig.~\ref{fig:PDF_Angle_Distribution} that the Fej\'er kernel function is decreasing monotonically (for increasing positive angles) within the support of the PDF of ordered $\theta_i$ and $\theta_j$ when $i\,{=}\,25$ and $j\,{=}\,20$. This means that the set of inequalities in the Fej\'er kernel based ordering of \eqref{eq:Fejer_ordering} can be equally represented by those in angle based ordering of \eqref{eq:angle_ordering}. In other words, the inequality ${\rm F}_M(\theta_j) \,{\geq}\, {\rm F}_M(\theta_i)$ always corresponds to $\theta_j \,{<}\, \theta_i$, and, hence, both these ordering schemes become equivalent. On the other hand, when we assume the user pair $i\,{=}\,50$ and $j\,{=}\,40$, the Fej\'er kernel function is non-monotonic within the support of the respective angle PDFs, i.e., it increases for $\theta_k \,{\leq}\, 0.02$~radian and decreases for $\theta_k \,{>}\, 0.02$~radian (along with increasing angle). Therefore, the inequalities in Fej\'er kernel based ordering do not necessarily match those in absolute angle based ordering. Or equivalently, the inequality ${\rm F}_M(\theta_j) \,{\geq}\, {\rm F}_M(\theta_i)$ corresponds to either $\theta_i \,{\geq}\, \theta_j$ or $\theta_i \,{<}\, \theta_j$ depending on the particular values of $\theta_i$ and $\theta_j$. As a result, sum rate performance of these two ordering strategy differs slightly when $i\,{=}\,50$ and $j\,{=}\,40$.

\begin{figure}[!t]
%\vspace{-1em}
\centering
\includegraphics[width=0.45\textwidth]{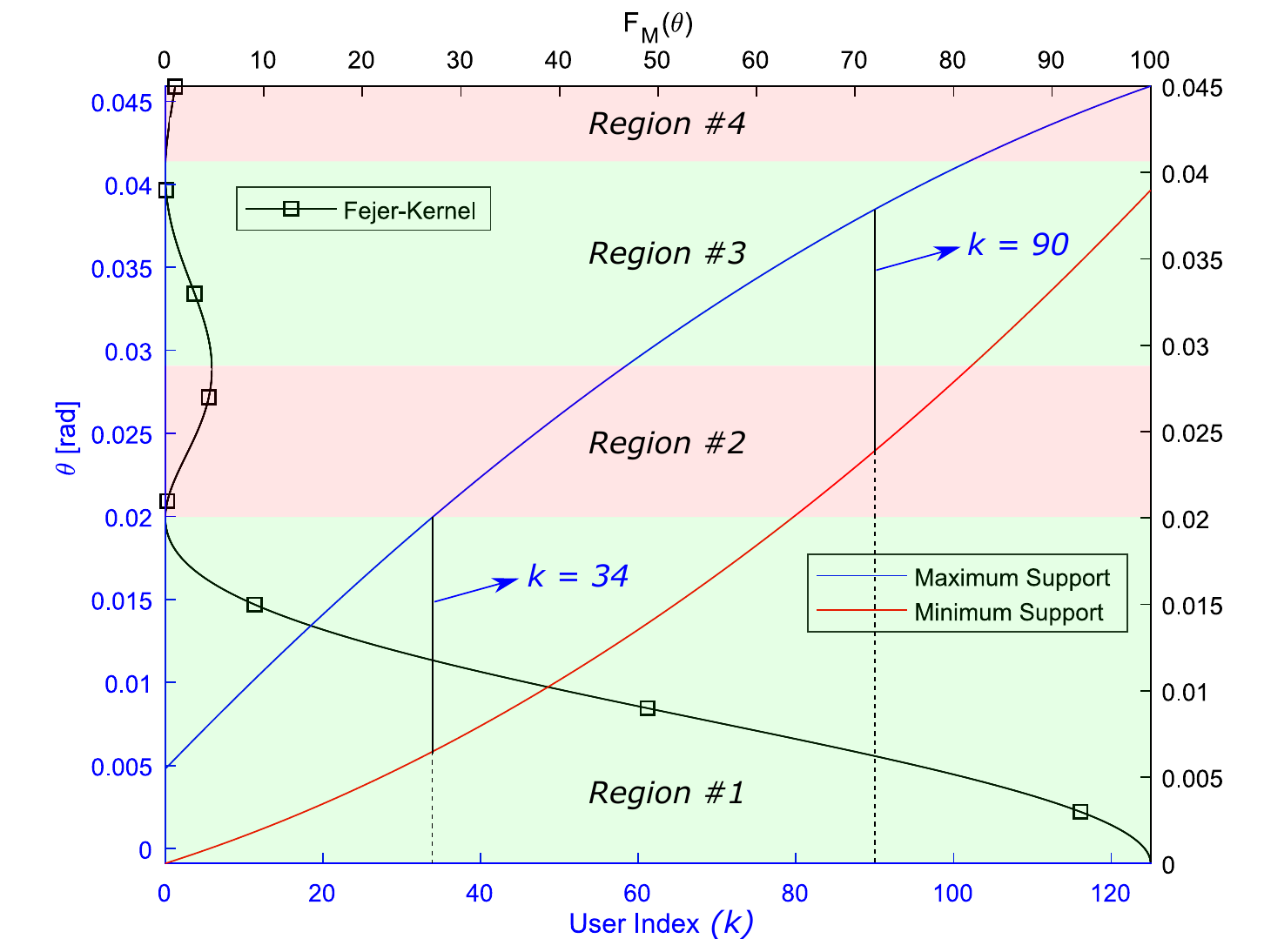}%\vspace{-1mm}
\caption{Support of angle PDF for the $k$-th user with $K\,{=}\,125$, $\Delta\,{=}\,5^{\circ}$ and angle based ordering. Entire $\theta$ support is partitioned into $4$ regions in which the Fej\'er kernel is changing monotonically.}
\label{fig:User_Idx_Angle_Distribution}
\end{figure}

In Fig.~\ref{fig:User_Idx_Angle_Distribution}, we further plot the minimum and maximum values of the support of the angle PDF, where the respective PDF is nonzero only within its support, for each user $k$ such that $1\,{\leq}\,k\,{\leq}\,125$. The users are assumed to be ordered according to angle based ordering given in~\eqref{eq:angle_ordering}. For each angle value $\theta$ (depicted along the vertical axis), we also plot the corresponding values of the Fej\'er kernel function (depicted along the upper horizontal axis). In addition, we split the entire $\theta$ range into four regions, each of which guarantees monotonic variation of the Fej\'er kernel function (i.e., either increasing or decreasing, but not both). As along as angle support of each NOMA user (i.e., both respective minimum and maximum support values) falls entirely into one of these four regions, respective ${\rm F}_M(\theta)$ is guaranteed to change monotonically. In such a case, the Fej\'er kernel and angle based ordering strategies become equivalent. As an example, the angle support of $k$-th users with $k\,{\leq}\,34$ completely lies within the $1$st region, where ${\rm F}_M(\theta)$ is a strictly decreasing function of $\theta$. However, the angle support of user $k\,{=}\,90$ lies within both the $2$nd and $3$rd regions, where ${\rm F}_M(\theta)$ is an increasing and decreasing function of $\theta$, respectively. Hence, as long as the NOMA user indices $i$ and $j$ are selected such that $j \,{<}\, i \,{\leq}\, 34$, ${\rm F}_M(\theta)$ changes monotonically, and both ordering strategies achieve the same outage sum rate performance, which agrees with the results presented in Fig.~\ref{fig:sumrate_angle_fejer_comparison_j20_i25}.

\subsection{Impact of User Region Geometry}

\begin{figure}[!t]
%\vspace{-2em}
\centering
%\hspace*{-0.6in}
\subfloat[Distance based ordering.]{\includegraphics[width=0.25\textwidth]{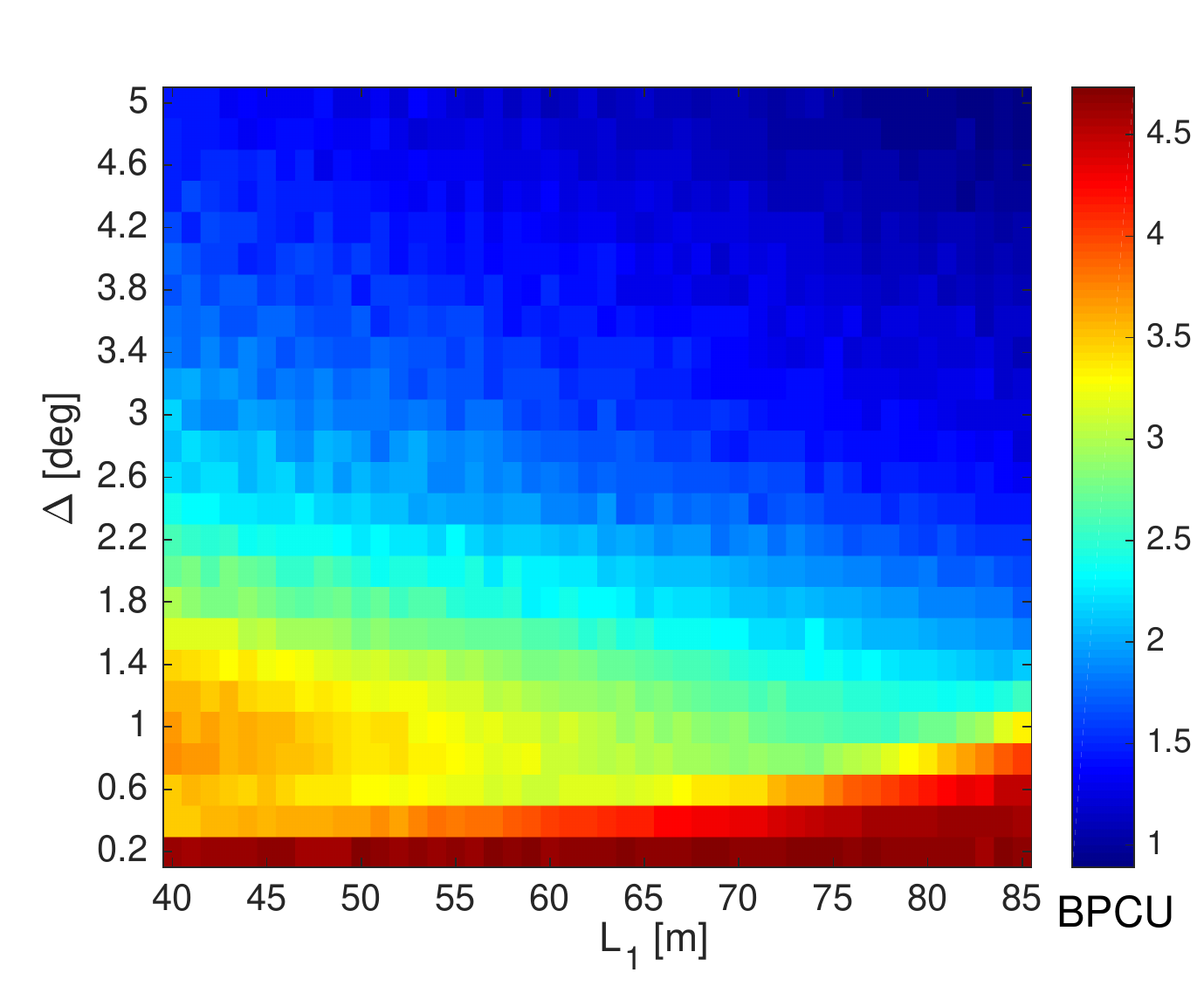}
\label{fig:Geometry_Distance_FB}}
\hspace*{-0.1in}
\subfloat[Fej\'er kernel based ordering.]{\includegraphics[width=0.25\textwidth]{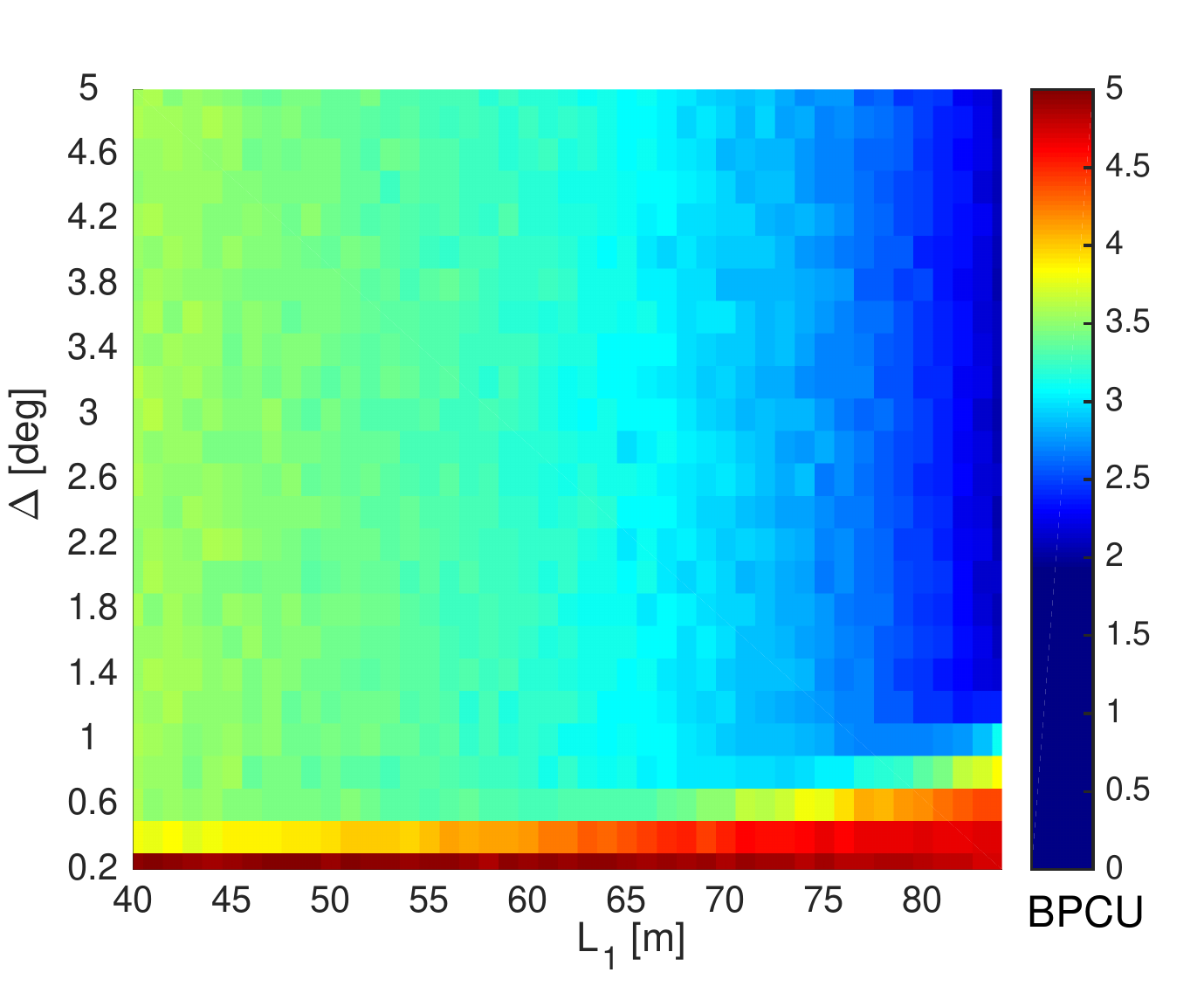}
\label{fig:Geometry_Fejer_Kernel_FB}}
\hspace*{-0.5in}\\
\subfloat[Rate difference = Distance - Fej\'er kernel.]{\includegraphics[width=0.45\textwidth]{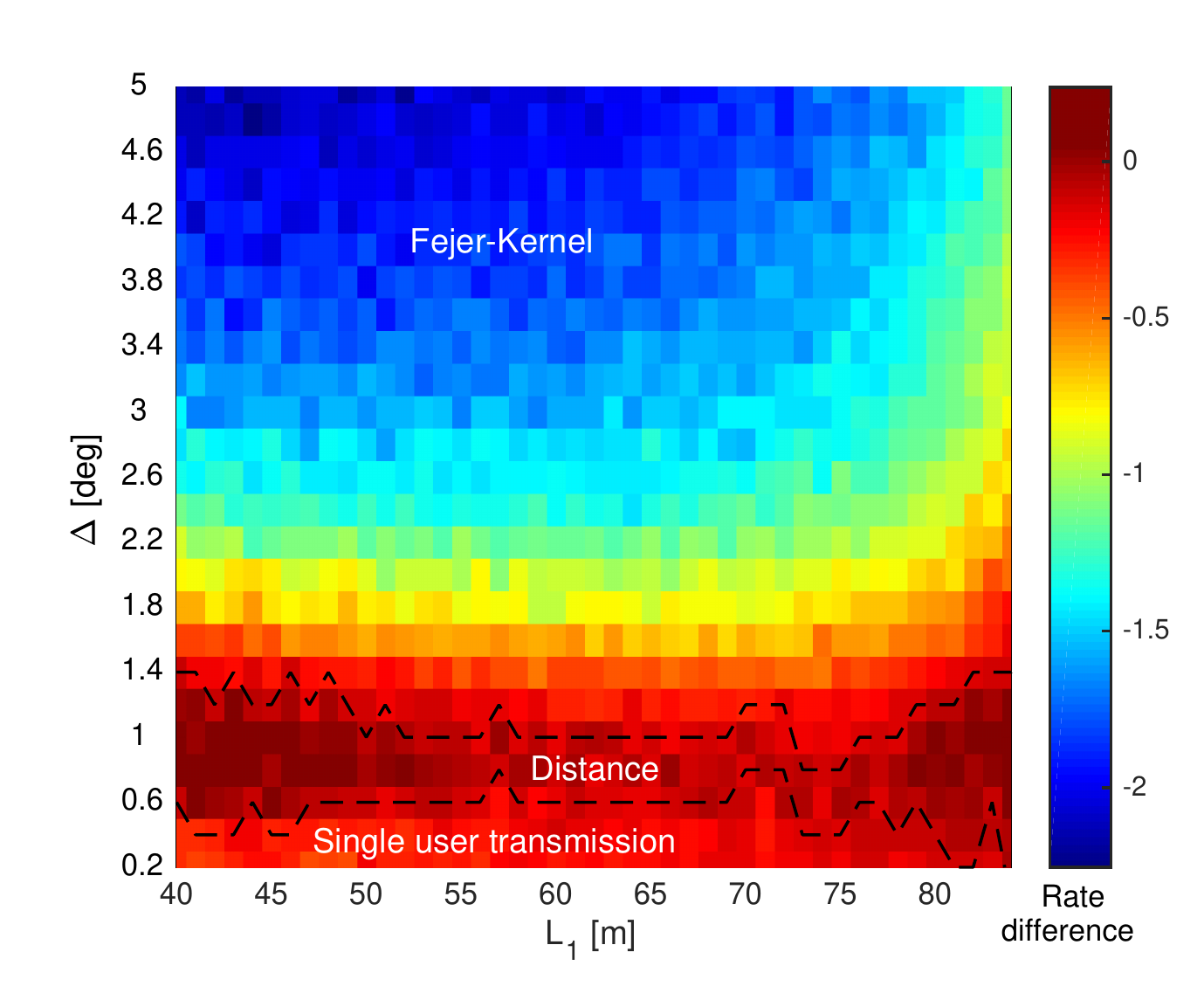}
\label{fig:Geometry_Distance_Fejer_Kernel_Rate_Difference}}
\caption{Outage sum rates for various user region geometries for distance and Fej\'er kernel based ordering together with their difference, where $i\,{=}\,25$, $j\,{=}\,20$, $P_{\rm Tx}\,{=}\,10$~dBm and $h\,{=}\,50$~m.}
\label{fig:Geometry_Fejer_Distance_FB}
\end{figure}

In this section, we study the impact of different user region geometries on the outage sum rate performance of distance and Fej\'er kernel based ordering strategies. In particular, we consider various user regions by letting $L_1\,{\in}\,[40,85]$~m (keeping $L_2$ the same) and $\Delta\,{\in}\,[0.2^\circ, 5^\circ]$ with $i\,{=}\,25$, $j\,{=}\,20$, $P_{\rm Tx}\,{=}\,10$~dBm, and $h\,{=}\,50$~m. The outage sum rate values are plotted in Fig.~\ref{fig:Geometry_Fejer_Distance_FB}\subref{fig:Geometry_Distance_FB} and Fig.~\ref{fig:Geometry_Fejer_Distance_FB}\subref{fig:Geometry_Fejer_Kernel_FB} for the distance and Fej\'er kernel ordering strategies, respectively, together with their difference in Fig.~\ref{fig:Geometry_Fejer_Distance_FB}\subref{fig:Geometry_Distance_Fejer_Kernel_Rate_Difference}.

We observe that when $\Delta$ is small, distance based ordering provides greater sum rates than that of Fej\'er kernel, since the NOMA users can be better distinguished based on their distance (not angle) information. On the other hand, as $\Delta$ gets larger, the NOMA users become more distinctive in angle (as compared to distance) and the Fej\'er kernel based ordering becomes superior. This behavior actually agrees with the results of the scenario considered in Section~\ref{sec:Perf_Rates_Dist_Fejer}.

Note that when we have $\Delta\,{\approx}\,0.2^{\circ}$, it is very unlikely to find both NOMA users in the user region, which basically depends on the particular choice of $i$ and $j$. In this case, there is either no transmission at all, or single user transmission may happen with scheduling $j$-th user all the time (i.e., $j \,{\leq}\, K \,{<}\, i$). Note also that higher sum rates can be expected whenever $j$-th user is selected employing Fej\'er kernel based ordering. This is because, the contribution of ${\rm F}_M(\theta)$ into the effective channel gain in \eqref{eq:Eff_channel_gain} is higher compared to the distance dependent PL term. As a result, difference in outage sum rates in Fig.~\ref{fig:Geometry_Fejer_Distance_FB}\subref{fig:Geometry_Distance_Fejer_Kernel_Rate_Difference} with the label ``single user transmission'' appear to be negative.

% \begin{figure}[!t]
% \centering
% \includegraphics[width=0.7\textwidth]{./img/Geometry_Distance_Fejer_Kernel_FB_10dBm_50m}\vspace{-1mm}
% \caption{Achievable rate difference between distance and Fej\'er kernel based user ordering considering different user region geometries, $i=25$, $j=20$, $P_{\rm Tx}=10$~dBm and $h=50$~m.}
% \label{fig:Geometry_Fejer_Distance_FB}
% \end{figure}

\section{Concluding Remarks}\label{sec:conclusion}
In this paper, we introduce NOMA transmission along with beamforming to a UAV-BS flying over a densely packed stadium providing broadband coverage. Beamforming allows NOMA to exploit space domain and as a result user angle information becomes a promising practical alternative to full CSI feedback for NOMA formulations. Considering user angle information we propose two user ordering strategies, 1) Fej\'er kernel based ordering, and 2) (absolute) angle based ordering.

% In addition, we consider user distance as another limited feedback scheme and investigate analytically and through extensive Monte carlo runs the achievable outage sum rate performance with these limited feedback schemes under various settings.

% Based on these feedback schemes, three user ordering strategies for NOMA formulation distance and angle (considering both ) during NOMA formulation.

% Our investigation shows that users should be ordered based on a channel quality measure (either distance or angle), on which users become more distinguishable.
Our investigation shows that when determining a feedback scheme as a measure of channel quality (either angle or distance), it is important to identify under which feedback scheme users become more distinguishable. For instance, whenever the footprint of the UAV beam on the ground is wide enough in horizontal angle, the outage sum rate performance of angle based ordering strategies outperform that of the distance based user ordering.
% On the other hand, if users become more distinctive on their respective distances, it is better to consider user distance as the limited feedback scheme.
Further, we show that whenever NOMA user pair has angle support over which the Fej\'er kernel function is monotonically varying, both Fej\'er kernel and angle based ordering strategies provide similar sum rate performance.

% we shed light on the behavior of two user ordering strategies with angle feedback scheme.  In particular, our investigation

% Even though the Fej\'er kernel based user ordering is the optimal strategy for angle feedback scheme, the performance of user ordering based on angle is also investigated rigorously.

\appendices

\begin{figure}[!t]
%\vspace{-1em}
\centering
%\hspace*{-0.3in}
\captionsetup[subfigure]{oneside,margin={0.5cm,0.5cm}}
\subfloat[Varying radial distance.]{\includegraphics[width=0.24\textwidth]{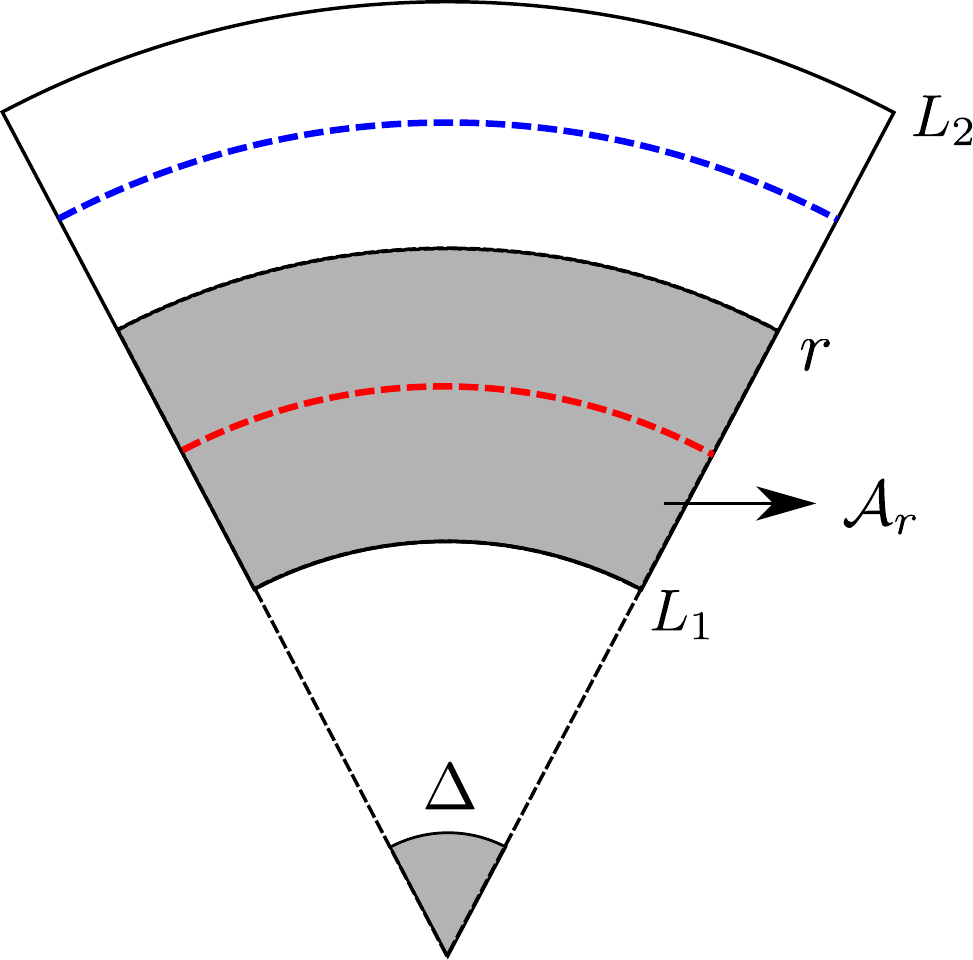}
\label{fig:user_region_radial_distance}}
\hspace*{0.1in}
\subfloat[Varying angle.]{\includegraphics[width=0.25\textwidth]{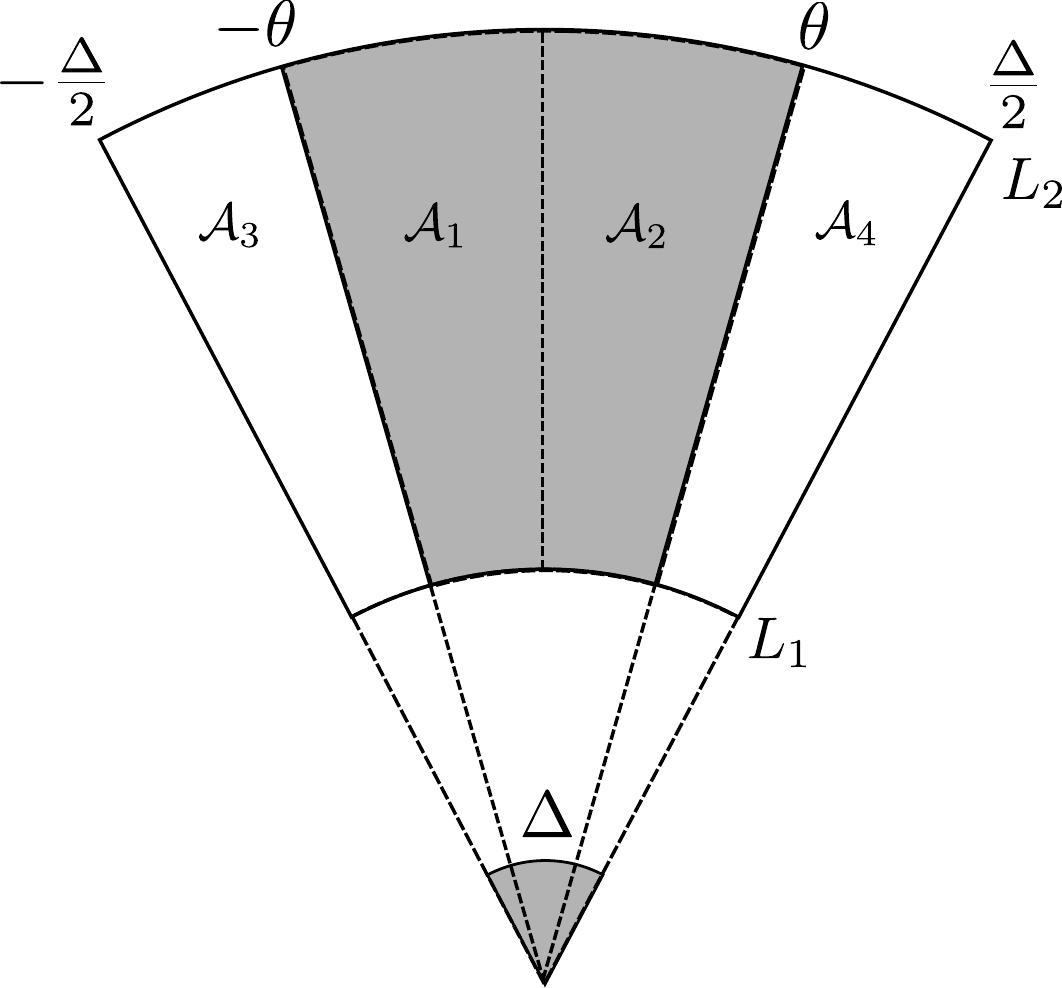}
\label{fig:user_region_angle}} \vspace{1em}
\caption{Horizontal footprint of the user region in Fig.~\ref{fig:footprint}.}
\label{fig:sketch_user_region}
\end{figure}

\section{The PDF of Unordered User Distance Distribution} \label{app:PDF_Unordered_user_dist}

In order to derive the unordered user distance distribution, we will relate the user distances to the number of users as discussed in \cite{Haenggi05Stochastic_Geo}. In Fig.~\ref{fig:sketch_user_region}, we sketch the projection of 3D scenario of Fig.~\ref{fig:footprint} on the horizontal plane. We observe in Fig.~\ref{fig:sketch_user_region}\subref{fig:user_region_radial_distance} that the number of users on each circular contour (having the same distance $r$ to the origin) increases with increasing $r$. For example, the red contour in Fig.~\ref{fig:sketch_user_region}\subref{fig:user_region_radial_distance} is likely to have less number of users as compared to that of blue contour, since red one is shorter in length than the blue one and users are uniformly distributed. We therefore intuitively conclude that the unordered distance distribution should take larger values with increasing $r$, which agrees with the respective PDF in \eqref{eq:PDF_unordered_distance}.

Specifically, the cumulative distribution function (CDF) of the unordered distance distribution can be written by considering average number of users in the area $\mathcal{A}_r$ as follows \vspace{-1em}

\footnotesize
\begin{align}\label{eq:CDF_Unorder__Dist_j_less_K_less_i}
F_d(r) = {\rm Pr}\{d \leq r\} = \frac{\mu\left( \mathcal{A}_r \right)}{\mu} = \frac{(r^2\,{-}\,L_1^2)\frac{\Delta}{2} \lambda}{(L_2^2\,{-}\,L_1^2)\frac{\Delta}{2} \lambda}=\frac{(r^2\,{-}\,L_1^2)}{(L_2^2\,{-}\,L_1^2)}.
\end{align} \normalsize
Taking the derivative of $F_d(r)$ we readily obtain \eqref{eq:PDF_unordered_distance}. \hfill\IEEEQEDhere

\section{The CDF of Unordered Fej\'er Kernel} \label{app:CDF_Unordered_Fejer_Kernel}

We start the derivation by splitting the Fej\'er Kernel function into small regions, where monotonic variation is guaranteed as shown in Fig.~\ref{fig:Fejer-Kernel_func}. These regions are defined as
\begin{align}\label{eq:regions_fejer_kernel}
\mathcal{R}_m: u=g_m(\theta) \textrm{ for } \{\theta_{m{-}1} \leq \theta \leq \theta_m\} ,
\end{align}
where $m \,{=}\, 1,2,3,4$, and $\theta_0 \,{=}\,0$. Since Fej\'er Kernel function is symmetric around $\theta \,{=}\, 0$, it is enough to consider only one side (i.e., either $\theta \,{<}\, 0$ or $\theta \,{\geq}\, 0$) while characterizing its distribution. Recalling that $U$ denotes unordered Fej\'er Kernel, the desired CDF is given as $F_U(u) \,{=}\, {\rm Pr}\{U \,{\leq}\, u\}$, which will be analyzed by considering the contribution of each region $\mathcal{R}_m$ separately for a specific set of Fej\'er Kernel values.

%Further, the CDF of unordered Fej\'er Kernel can be derived considering the angle distribution $\theta$. In particular, we know that $\theta$ is distributed uniformly randomly. Hence, by looking at the corresponding angle(s) for a given Fej\'er Kernel value $u$, it is possible to determine necessary angle probabilities to characterize Fej\'er Kernel distribution.

Let us first consider the case where the Fej\'er Kernel value $u$ satisfies ${\rm F}_M(\theta_1) \,{\leq}\, u \,{\leq}\, 100$. We observe from Fig.~\ref{fig:Fejer-Kernel_func} that the respective angle values $g_1^{-1}(u)$ (which generates specified $u$ values) are within the region $\mathcal{R}_1$ only. As a result, the desired CDF for ${\rm F}_M(\theta_1) \,{\leq}\, u \,{\leq}\, 100$ is given as \vspace{-1em}

\small
\begin{align} \label{eq:CDF_Fejer_part1}
F_U(u) = {\rm Pr}\{U \,{\leq}\, u\}={\rm Pr}\{\theta \geq g_1^{-1}(u)\}=1-\frac{g_1^{-1}(u)}{\Delta/2}.
\end{align} \normalsize

Similarly, when $u$ is such that ${\rm F}_M(\theta_2) \,{\leq}\, u \,{\leq}\, {\rm F}_M(\theta_1)$, there are three possible angle values $g_1^{-1}(u)$, $g_2^{-1}(u)$, and $g_3^{-1}(u)$ producing this specific the Fej\'er Kernel value $u$, and these angles lie within regions $\mathcal{R}_1$, $\mathcal{R}_2$, and $\mathcal{R}_3$, respectively. The desired CDF for ${\rm F}_M(\theta_2) \,{\leq}\, u \,{\leq}\, {\rm F}_M(\theta_1)$ becomes \vspace{-1em}

\small
\begin{align} \label{eq:CDF_Fejer_part2}
F_U(u)&= {\rm Pr}\{U \,{\leq}\, u\}\nonumber \\
&={\rm Pr}\{g_1^{-1}(u) \leq \theta \leq g_2^{-1}(u)\}+{\rm Pr}\{g_3^{-1}(u) \leq \theta \leq \theta_4 \}\nonumber \\
&=\frac{g_2^{-1}(u)-g_1^{-1}(u)}{\Delta/2} + \frac{\theta_4-g_3^{-1}(u)}{\Delta/2}.
\end{align} \normalsize

Finally, whenever we have $u \leq {\rm F}_M(\theta_2)$, there are four corresponding angles $g_1^{-1}(u)$, $g_2^{-1}(u)$, $g_3^{-1}(u)$, and $g_4^{-1}(u)$ lying within $\mathcal{R}_1$, $\mathcal{R}_2$, $\mathcal{R}_3$, and $\mathcal{R}_4$, respectively. The CDF for this particular condition $u \leq {\rm F}_M(\theta_2)$ is given as \vspace{-1em}

\small
\begin{align} \label{eq:CDF_Fejer_part3}
&F_U(u)= {\rm Pr}\{U \,{\leq}\, u \} \nonumber \\
&={\rm Pr}\{g_1^{-1}(u) \leq \theta \leq g_2^{-1}(u)\}+{\rm Pr}\{g_3^{-1}(u) \leq \theta \leq g_4^{-1}(u) \}\nonumber \\
&=\frac{g_2^{-1}(u)-g_1^{-1}(u)}{\Delta/2} + \frac{g_4^{-1}(u)-g_3^{-1}(u)}{\Delta/2},
\end{align} \normalsize
which yields \eqref{eq:CDF_unordered_Fejer_Kernel} together with \eqref{eq:CDF_Fejer_part1} and \eqref{eq:CDF_Fejer_part2}. \hfill  \IEEEQEDhere

\section{The PDF of Ordered $k$-th User Absolute Angle} \label{app:PDF_Ordered_Abs_Angle}

Since we need to have $K\,{\geq}\,j$ to start transmission, we derive absolute angle distribution for $\mathcal{S}_{K_{2}}{:}\left\lbrace K \,|\, j\,{<}\, K \,{\leq}\, i \right\rbrace$ and $\mathcal{S}_{K_3}{:}\left\lbrace K  \,|\, K \,{\geq}\, i \right\rbrace$, separately. We first consider the CDF of the $k$-th user absolute angle $\tilde{\theta}_k$ for $\mathcal{S}_{K_2}$ and $\bar{\theta}\,{=}\,0$, and hence $\tilde{\theta}_k\,{=}\,|\theta_k|$. The CDF $F_{\tilde{\theta}_k|\mathcal{S}_{K_2}} (\theta)$ is defined as \vspace{-1em}

\footnotesize
\begin{align} \label{eq:CDF_j_less_K_less_i}
F_{\tilde{\theta}_k|\mathcal{S}_{K_2}} (\theta) = {\rm Pr}\{\tilde{\theta}_k\,{\leq}\,\theta | \, j \,{\leq}\, K \,{<}\, i \} &= \frac{{\rm Pr}\{ \tilde{\theta}_k\,{\leq}\, \theta, \ j \,{\leq}\, K \,{<}\, i \} }{{\rm Pr}\{ j \,{\leq}\, K \,{<}\, i \} }.
% \\ \nonumber
% & \hspace{10em} + \frac{{\rm P}\{ \tilde{\theta}_k\,{<}\,-\theta, \theta<0 , \ j \,{\leq}\, K \,{<}\, i \} }{{\rm P}\{ j \,{\leq}\, K \,{<}\, i \} }.
\end{align} \normalsize
The denominator of \eqref{eq:CDF_j_less_K_less_i} is readily available from the definition of HPPP, while we need to derive the probability in the numerator. For this purpose, we relate the ordered absolute angles to the number of users by making use of the geometry in Fig.~\ref{fig:sketch_user_region}\subref{fig:user_region_angle}. We assume that $\theta$ takes any value within the range $0\,{\leq}\,\theta\,{\leq}\,\frac{\Delta}{2}$, and hence the entire angular range $\left[{-}\frac{\Delta}{2},\frac{\Delta}{2}\right]$ is spanned by this definition. For the condition $\tilde{\theta}_k\,{\leq}\,\theta$ to be satisfied, it is necessary that the area $\mathcal{A}_1\,{\cup}\,\mathcal{A}_2$ in Fig.~\ref{fig:sketch_user_region}\subref{fig:user_region_angle} has at least $k$ users. In addition, for a given number of users $l\,{\geq}\,k$ within $\mathcal{A}_1\,{\cup}\,\mathcal{A}_2$, the remaining area $\mathcal{A}_3\,{\cup}\,\mathcal{A}_4$ should have at most $(i\,{-}\,l\,{-}\,1)$ users to satisfy $\mathcal{S}_{K_2}$ (i.e., $j \,{\leq}\, K \,{<}\, i$). Equivalently, this implies that the user region should have less than $i$ users. Hence, by considering $L\,{=}\,(L_2^2{-}L_1^2)\lambda$, the desired probability is calculated as  \vspace{-1em}

\small
\begin{align}
&{\rm Pr}\{ \tilde{\theta}_k\,{\leq}\, \theta, \ j \,{\leq}\, K \,{<}\, i \} \,{=} \nonumber \\
& \sum \limits_{l=k}^{i-1} {\rm Pr}  \{\mathcal{A}_1\,{\cup}\,\mathcal{A}_2 \, \textrm{has $l$ users}, \, \mathcal{A}_3\,{\cup}\,\mathcal{A}_4 \, \textrm{has at most $(i\,{-}\,l\,{-}\,1)$ users}\} \nonumber \\
&= \sum \limits_{l=k}^{i{-}1} \frac{e^{{-}\theta L}\left[ \theta L \right]^l}{l!}\left\lbrace \sum \limits_{l'=0}^{i{-}l{-}1} \frac{e^{{-}(\frac{\Delta}{2}-\theta)L}\left[ (\frac{\Delta}{2}-\theta)L \right]^{l'}}{l'!}\right\rbrace. \label{eq:marginal_cdf}
\end{align} \normalsize
Employing $\mathcal{C}\,{=}\, {\rm Pr}\{ j \,{\leq}\, K \,{<}\, i \} \,{=}\, \sum\limits_{l=j}^{i{-}1} \frac{e^{{-}\frac{\Delta}{2}(L_2^2{-}L_1^2)\lambda} \left[\frac{\Delta}{2}(L_2^2{-}L_1^2)\lambda \right]^l}{l!}$ and \eqref{eq:marginal_cdf}, the CDF in \eqref{eq:CDF_j_less_K_less_i} is readily available while the respective PDF is obtained as \vspace{-1em}

\footnotesize
\begin{align} \label{eq:PDF_ordered_abs_angle_Sk2}
f_{\tilde{\theta}_k|\mathcal{S}_{K_2}} (\theta)&=\frac{e^{-\frac{\Delta}{2} L}}{\mathcal{C}}\frac{{\rm d} }{{\rm d} \theta} \left\lbrace  \sum \limits_{l=k}^{i-1} \sum \limits_{l'=0}^{i-l-1} \frac{\left[ \theta L \right]^l}{l!} \frac{\left[ (\frac{\Delta}{2}- \theta) L \right]^{l'}}{l'!}\right\rbrace \nonumber \\
&=\frac{ L}{\mathcal{C}}e^{-\frac{\Delta}{2} L}\frac{\left[\theta L \right]^{(k-1)}}{(k-1)!} \Bigg( \sum \limits_{l=0}^{i-k-1}\frac{\left[(\frac{\Delta}{2}-\theta)L \right]^{l}}{l!} \Bigg). %\qquad \; \IEEEQEDhere
\end{align} \normalsize
% \section{The Joint PDF of User Distances for $\mathcal{S}_{K_2}$} \label{app:JPDF_K_greater_i}
Similarly, the CDF of $k$-th user absolute angle $\tilde{\theta}_k$ for $\mathcal{S}_{K_3}$ is
\begin{align}\label{eq:K_greater_i}
F_{\tilde{\theta}_k|\mathcal{S}_{K_3}} (\theta) \,{=}\, {\rm Pr}\{\tilde{\theta}_k\,{\leq}\,\theta | \, K \,{\geq}\, i \}  \,{=}\, \frac{{\rm Pr}\{\tilde{\theta}_k\,{\leq}\,\theta, K \geq i \} }{{\rm Pr}\{ K \geq i \} }.
\end{align}
Considering a similar argument as in the case for $\mathcal{S}_{K_2}$, the probability in the numerator of \eqref{eq:K_greater_i} can be derived by relating that to the number of users in the area. In particular, $\mathcal{A}_1\,{\cup}\,\mathcal{A}_2$ in Fig.~\ref{fig:sketch_user_region}\subref{fig:user_region_angle} should have at least $k$ users to satisfy $\tilde{\theta}_k\,{\leq}\,\theta$. Given that there are $l$ users with $k\,{\leq}\, l\,{<}\,i$ in $\mathcal{A}_1\,{\cup}\,\mathcal{A}_2$, we need to have at least $i\,{-}\,l$ users in the area $\mathcal{A}_3\,{\cup}\,\mathcal{A}_4$ to satisfy $\mathcal{S}_{K_3}$ (i.e., $K \,{\geq}\, i$). On the other hand, if $\mathcal{A}_1\,{\cup}\,\mathcal{A}_2$ has more than $i$ users, both conditions in the numerator of \eqref{eq:K_greater_i} is satisfied, and hence the desired probability can be derived as follows \vspace{-2em}

\small
\begin{align}
&{\rm Pr}\{\tilde{\theta}_k\,{\leq}\,\theta, K \geq i \} = \sum \limits_{l=k}^{i-1} {\rm Pr}  \{\mathcal{A}_1\,{\cup}\,\mathcal{A}_2 \, \textrm{has $l$ users}, \nonumber \\
&\, \mathcal{A}_3\,{\cup}\,\mathcal{A}_4 \, \textrm{has at least $(i-l)$ users}\}+\sum \limits_{l=i}^{\infty} {\rm Pr}  \{\mathcal{A}_1\,{\cup}\,\mathcal{A}_2 \, \textrm{has $l$ users} \}, \nonumber \\
& = \sum \limits_{l=k}^{i-1} \frac{e^{-\theta L}\left[ \theta L \right]^l}{l!}\left\lbrace 1 - \sum \limits_{l'=0}^{i-l-1} \frac{e^{-(\frac{\Delta}{2} -\theta) L}\left[(\frac{\Delta}{2} -\theta) L \right]^{l'}}{l'!}\right\rbrace \nonumber \\
&\hspace{13em}+ \left\lbrace 1 - \sum \limits_{l=0}^{i-1} \frac{e^{-\theta L}\left[ \theta L\right]^{l}}{l!}\right\rbrace \label{eq:CDF_proof_step_3}.
\end{align} \normalsize
Employing \eqref{eq:CDF_proof_step_3} and $\mathcal{C} \,{=}\, {\rm Pr}\{ K \geq i \} \,{=}\, 1 - \sum \limits_{l = 0}^{i-1} \frac{e^{-\frac{\Delta}{2} L} \left[\frac{\Delta}{2} L \right]^l}{l!}$, the CDF in \eqref{eq:K_greater_i} is readily available. Taking the derivative of the CDF, we obtain the PDF for $\mathcal{S}_{K_3}$ as in \eqref{eq:PDF_ordered_abs_angle_Sk3}.\hfill\IEEEQEDhere

% \small
% \begin{align} \label{eq:PDF_ordered_abs_angle_Sk3}
% &f_{\tilde{\theta}_k|\mathcal{S}_{K_3}} (\theta) = \frac{{\rm d}}{{\rm d} \theta } F_{\tilde{\theta}_k|\mathcal{S}_{K_3}} (\theta) \nonumber \\
% &= \frac{1}{\mathcal{C}} \frac{{\rm d}}{{\rm d} \theta }  \left\lbrace \sum \limits_{l=k}^{i-1} e^{-\theta L} \frac{\left[ \theta L \right]^l}{l!} \right\rbrace - \frac{e^{-\frac{\Delta}{2} L}}{\mathcal{C}}\frac{{\rm d}}{{\rm d} \theta}  \left\lbrace \sum \limits_{l=k}^{i-1} \sum \limits_{l'=0}^{i-l-1} \frac{\left[ \theta L \right]^l}{l!} \frac{\left[ (\frac{\Delta}{2}-\theta) L \right]^{l'}}{l'!}\right\rbrace+\frac{1}{\mathcal{C}} \frac{{\rm d}}{{\rm d} \theta }  \left\lbrace 1- \sum \limits_{l=0}^{i-1} e^{-\theta L} \frac{\left[ \theta L \right]^l}{l!} \right\rbrace
% \nonumber \\
% &= \frac{Le^{-\theta L}}{\mathcal{C}} \left\lbrace  \frac{ \left[ \theta L \right]^{(k-1)}}{(k-1)!} - \frac{ \left[ \theta L \right]^{(i-1)}}{(i-1)!}  \right\rbrace - \frac{Le^{-\frac{\Delta}{2} L}}{\mathcal{C}}  \frac{ \left[ \theta L \right]^{(k-1)}}{(k-1)!} \left\lbrace \sum \limits_{l'=0}^{i-k-1} \frac{\left[ (\frac{\Delta}{2}-\theta)L \right]^{l'}}{l'!} \right\rbrace+\frac{Le^{-\theta L}}{\mathcal{C}} \left\lbrace  \frac{ \left[ \theta L \right]^{(i-1)}}{(i-1)!}  \right\rbrace
% \nonumber \\
% &=\frac{ L}{\mathcal{C}}\frac{\left[\theta L \right]^{(k-1)}}{(k-1)!} \left\lbrace e^{-\theta L}- e^{-\frac{\Delta}{2} L} \sum \limits_{l=0}^{i-k-1}\frac{\left[(\frac{\Delta}{2}-\theta) L \right]^{l}}{l!} \right\rbrace. \hspace{19.5em} \IEEEQEDhere
% \end{align}\normalsize

\begin{figure*}[!htb]
\setcounter{MYtempeqncnt}{\value{equation}}
\setcounter{equation}{43} \footnotesize
\begin{align} \label{eq:PDF_ordered_abs_angle_Sk3}
f_{\tilde{\theta}_k|\mathcal{S}_{K_3}} (\theta) &= \frac{{\rm d}}{{\rm d} \theta } F_{\tilde{\theta}_k|\mathcal{S}_{K_3}} (\theta) = \frac{1}{\mathcal{C}} \frac{{\rm d}}{{\rm d} \theta }  \left\lbrace \sum \limits_{l=k}^{i-1} e^{-\theta L} \frac{\left[ \theta L \right]^l}{l!} \right\rbrace - \frac{e^{-\frac{\Delta}{2} L}}{\mathcal{C}}\frac{{\rm d}}{{\rm d} \theta}  \left\lbrace \sum \limits_{l=k}^{i-1} \sum \limits_{l'=0}^{i-l-1} \frac{\left[ \theta L \right]^l}{l!} \frac{\left[ (\frac{\Delta}{2}-\theta) L \right]^{l'}}{l'!}\right\rbrace+\frac{1}{\mathcal{C}} \frac{{\rm d}}{{\rm d} \theta }  \left\lbrace 1- \sum \limits_{l=0}^{i-1} e^{-\theta L} \frac{\left[ \theta L \right]^l}{l!} \right\rbrace
\nonumber \\
&= \frac{Le^{-\theta L}}{\mathcal{C}} \left\lbrace  \frac{ \left[ \theta L \right]^{(k-1)}}{(k-1)!} - \frac{ \left[ \theta L \right]^{(i-1)}}{(i-1)!}  \right\rbrace - \frac{Le^{-\frac{\Delta}{2} L}}{\mathcal{C}}  \frac{ \left[ \theta L \right]^{(k-1)}}{(k-1)!} \left\lbrace \sum \limits_{l'=0}^{i-k-1} \frac{\left[ (\frac{\Delta}{2}-\theta)L \right]^{l'}}{l'!} \right\rbrace+\frac{Le^{-\theta L}}{\mathcal{C}} \left\lbrace  \frac{ \left[ \theta L \right]^{(i-1)}}{(i-1)!}  \right\rbrace
\nonumber \\
&=\frac{ L}{\mathcal{C}}\frac{\left[\theta L \right]^{(k-1)}}{(k-1)!} \left\lbrace e^{-\theta L}- e^{-\frac{\Delta}{2} L} \sum \limits_{l=0}^{i-k-1}\frac{\left[(\frac{\Delta}{2}-\theta) L \right]^{l}}{l!} \right\rbrace.
\end{align} \normalsize
\setcounter{equation}{44}
\hrulefill
\end{figure*}

\section{The PDF of Ordered $k$-th User Angle Distribution} \label{app:PDF_Ordered_Angle}

In this section, we derive the PDF of the ordered $k$-th user angle distribution considering ordered the $k$-th user absolute angle distribution derived in Appendix~\ref{app:PDF_Ordered_Abs_Angle}. The CDF of the ordered $k$-th user angle, $F_{{\theta}_k|\mathcal{S}_{K_m}} (\theta)$ with $m\in\{2,\ 3\}$ is
\begin{align} \label{eq:CDF_Ordered_Angle}
F_{{\theta}_k|\mathcal{S}_{K_m}} (\theta)={\rm Pr} \{{\theta}_k \leq \theta |\mathcal{S}_{K_m} \}.
% = \frac{1}{2}{\rm P} \{{\theta}_k < \theta,\ \theta \geq 0 |\mathcal{S}_{K_m}\}  + \frac{1}{2}{\rm P} \{{\theta}_k < \theta,\ \theta < 0|\mathcal{S}_{K_m} \},
\end{align}
% Note here that, since $\theta$ is uniformly distributed between $-\Delta$ and $\Delta$, ${\rm P} \{\theta \geq 0\}={\rm P} \{\theta < 0\}=\frac{1}{2}$.
where $\theta\,{\in}\,\left[{-}\Delta,\Delta\right]$. In the following, we consider the desired CDF for $\theta \,{\geq}\, 0$ and $\theta \,{<}\, 0$, separately. We first consider \eqref{eq:CDF_Ordered_Angle} for $\theta \,{\geq}\, 0$. Considering Fig.~\ref{fig:sketch_user_region}\subref{fig:user_region_angle}, the $k$-th user might be in $\mathcal{A}_1$, $\mathcal{A}_2$, or $\mathcal{A}_3$ in order to satisfy $\left\lbrace {\theta}_k \,{\leq}\, \theta,\theta \,{\geq}\, 0 \right\rbrace$ for a given $\mathcal{S}_{K_m}$. The range of values ${\theta}_k$ can take when the $k$-th user is in $\mathcal{A}_1$ or $\mathcal{A}_2$, respectively, is given for absolute angle $\tilde{\theta}_k$ as follows
\begin{align}
  \begin{array}{l@{}l}
		\mathcal{A}_1:{-}&{\theta} \leq {\theta}_k \leq 0 \,, 	\\
		\mathcal{A}_2: &0 \leq {\theta}_k \leq \theta \,,
  \end{array}
\end{align}
for $0 \,{\leq}\, \tilde{\theta}_k \,{\leq}\, \theta$. Similarly, if the $k$-th user is in $\mathcal{A}_3$ or $\mathcal{A}_4$, the range of ${\theta}_k$ values can be represented jointly as $\theta \,{\leq}\, \tilde{\theta}_k \,{\leq}\, \frac{\Delta}{2}$. With these definitions, ${\rm Pr} \{{\theta}_k \,{\leq}\, \theta, \theta \,{\geq}\, 0 |\mathcal{S}_{K_m}\}$ is given as \vspace{-1em}

\footnotesize
\begin{align} \label{eq:CDF_Angle_theta_greater_0}
{\rm Pr} \{{\theta}_k \,{\leq}\, \theta, \theta \geq 0 |\mathcal{S}_{K_m}\} &= {\rm Pr} \{\tilde{\theta}_k \,{\leq}\, \theta |\mathcal{S}_{K_m} \} + \frac{1}{2}\left\lbrace 1 - {\rm Pr} \{\tilde{\theta}_k \,{\leq}\,\theta |\mathcal{S}_{K_m} \} \right\rbrace \nonumber\\
&= \frac{1}{2} \left( 1+ {\rm P} \{\tilde{\theta}_k \,{\leq}\,\theta |\mathcal{S}_{K_m} \} \right).
\end{align} \normalsize
Note that, either $\mathcal{A}_3$ or $\mathcal{A}_4$ contributes equally in the probability term ${\rm P} \{\theta \leq \tilde{\theta}_k \leq \frac{\Delta}{2} |\mathcal{S}_{K_m} \}$.

We follow a similar strategy to determine ${\rm Pr} \{{\theta}_k \,{\leq}\, \theta, \theta \,{<}\, 0 |\mathcal{S}_{K_m}\}$ for $\theta\,{<}\,0$. In particular, the $k$-th user should be within $\mathcal{A}_3$ to satisfy $\left\lbrace {\theta}_k \,{\leq}\, \theta,\theta \,{<}\, 0 \right\rbrace$ for a given $\mathcal{S}_{K_m}$. Then the range of values ${\theta}_k$ can take (when the $k$-th user is in $\mathcal{A}_3$ or $\mathcal{A}_4$) is given as follows
\begin{align}
 \left.
  \begin{array}{r@{}l}
		\mathcal{A}_3:-\frac{\Delta}{2} \leq {\theta}_k \leq \theta \,, 	\\
		\mathcal{A}_4: -\theta \leq {\theta}_k \leq \frac{\Delta}{2} \,,
  \end{array}
  \right.
\end{align}
for ${-}\theta \,{\leq}\, \tilde{\theta}_k \,{\leq}\, \frac{\Delta}{2}$. Hence, the desired probability is given as \small
\begin{align} \label{eq:CDF_Angle_theta_less_0}
{\rm Pr} \{{\theta}_k\,{\leq}\, \theta, \theta < 0 |\mathcal{S}_{K_m}\}&= 0.5\Big( 1 - {\rm Pr} \{\tilde{\theta}_k \,{\leq}\, -\theta |\mathcal{S}_{K_m} \} \Big) \nonumber \\
&= 0.5 \Big( 1-{\rm Pr} \{\tilde{\theta}_k \,{\leq}\, -\theta |\mathcal{S}_{K_m} \} \Big).
\end{align} \normalsize
Using \eqref{eq:CDF_Angle_theta_greater_0} and \eqref{eq:CDF_Angle_theta_less_0} in \eqref{eq:CDF_Ordered_Angle}, CDF $F_{{\theta}_k|\mathcal{S}_{K_m}} (\theta)$ is given as \vspace{-1em}

\small
\begin{align}
F_{{\theta}_k|\mathcal{S}_{K_m}} (\theta)& ={\rm Pr} \{{\theta}_k \,{\leq}\,\theta |\mathcal{S}_{K_m} \}
\nonumber \\
&=\begin{cases}
\displaystyle 0.5 \Big( 1+{\rm Pr} \{\tilde{\theta}_k \,{\leq}\, \theta |\mathcal{S}_{K_m} \} \Big), \hfill \theta \geq 0, \\
\displaystyle 0.5 \Big( 1-{\rm Pr} \{\tilde{\theta}_k\,{\leq}\,-\theta |\mathcal{S}_{K_m} \} \Big), \hfill  \theta < 0,
\end{cases} \nonumber \\
&= 0.5 \Big( 1+{\rm sgn}(\theta) \, {\rm Pr} \left\lbrace \tilde{\theta}_k \,{\leq}\,|\theta| \mid \mathcal{S}_{K_m} \right\rbrace \Big), \label{eq:CDF_Ordered_Angle_1}
\end{align} \normalsize
where ${\rm sgn}(\cdot)$ denotes the signum function. Taking the derivative of $F_{{\theta}_k|\mathcal{S}_{K_m}} (\theta)$ produces the respective PDF as follows \vspace{-2em}

\small
 \begin{align}
f_{{\theta}_k|\mathcal{S}_{K_m}} (\theta) &= \frac{\rm d}{{\rm d}\theta}F_{{\theta}_k|\mathcal{S}_{K_m}} (\theta) \nonumber \\
&= \begin{cases}
\displaystyle 0.5 \frac{\rm d}{{\rm d}\theta}{\rm Pr} \{\tilde{\theta}_k \,{\leq}\,\theta |\mathcal{S}_{K_m} \}, \hfill  \theta \geq 0 ,\\
\displaystyle -0.5 \frac{\rm d}{{\rm d}\theta}{\rm Pr} \{\tilde{\theta}_k \,{\leq}\, -\theta |\mathcal{S}_{K_m} \}, \hfill \theta < 0 ,
\end{cases} \\
&= \frac{{\rm sgn}(\theta)}{2}\frac{\rm d}{{\rm d}\theta}{\rm Pr} \left\lbrace \tilde{\theta}_k \,{\leq}\,|\theta| \mid \mathcal{S}_{K_m} \right\rbrace . \label{eq:PDF_Ordered_Angle_1}
\end{align} \normalsize
Note that, $f_{\tilde{\theta}_k|\mathcal{S}_{K_m}} (\theta) \,{=}\, \frac{\rm d}{{\rm d}\theta}{\rm Pr} \{\tilde{\theta}_k \,{\leq}\, \theta |\mathcal{S}_{K_m} \}$ as derived in \eqref{eq:PDF_ordered_abs_angle_Sk2} and \eqref{eq:PDF_ordered_abs_angle_Sk3}. Furthermore, following the steps in Appendix~\ref{app:PDF_Ordered_Abs_Angle} with $\theta \,{<}\, 0$, we readily obtain ${-}f_{\tilde{\theta}_k|\mathcal{S}_{K_m}} (|\theta|) \,{=}\, \frac{\rm d}{{\rm d}\theta}{\rm Pr} \{\tilde{\theta}_k \,{\leq}\, {-}\theta |\mathcal{S}_{K_m} \}$. Hence, the PDF in \eqref{eq:PDF_Ordered_Angle_1} can be given for $m \in\, \{2,\ 3\}$ as follows
\begin{equation}
f_{{\theta}_k|\mathcal{S}_{K_m}} (\theta) =\frac{1}{2}f_{\tilde{\theta}_k|\mathcal{S}_{K_m}}(|\theta|). \IEEEQEDhereeqn\quad
\end{equation}

\section{The PDF of Ordered $k$-th User Distance for $\mathcal{S}_{K_3}$} \label{app:PDF_Ordered_Distance_K_greater_i}
Let us first consider the cumulative distribution function (CDF) of the $k$-th user distance $d_k$ assuming that $K$ is chosen from $\mathcal{S}_{K_3}{:} \{ K | K \,{\geq}\, i\}$. This can be given as \vspace{-1em}

\small
\begin{align}\label{eq:CDF_K_greater_i}
F_{d_k|\mathcal{S}_{K_3}} (r) = {\rm Pr}\{d_k \,{\leq}\, r | \, K \,{\geq}\, i\} = \frac{{\rm Pr}\{ d_k < r, \ K \,{\geq}\, i \} }{{\rm Pr}\{ K \,{\geq}\, i \} }.
\end{align} \normalsize
Note that while the denominator of \eqref{eq:CDF_K_greater_i} is readily available from the definition of HPPP, we will relate the ordered user distances to the number of users with the help of Fig.~\ref{fig:sketch_user_region}\subref{fig:user_region_radial_distance} in order to derive the probability in the numerator. To this end, the first condition $d_k\,{\leq}\,r$ in the numerator of \eqref{eq:CDF_K_greater_i} is interpreted as the necessity of the area $\mathcal{A}_r$ having at least $k$ users. Given that there are $l$ users with $k\,{\leq}\, l\,{<}\,i$ in $\mathcal{A}_r$, we need to have at least $(i\,{-}\,l)$ users in the rest of the area in user region to satisfy $\mathcal{S}_{K_3}$ (i.e., $K \,{\geq}\, i$). On the other hand, if $\mathcal{A}_r$ has more than $i$ users, both conditions in the numerator of \eqref{eq:CDF_K_greater_i} is satisfied. Letting  $\mathcal{A} \,{=}\, \frac{\Delta}{2}(L_2^2 \,{-}\, L_1^2)$, the desired probability can be derived as \vspace{-1em}

\small
\begin{align}
&{\rm Pr}\{d_k\,{\leq}\,r, K \geq i \} = \sum \limits_{l=k}^{i-1} {\rm Pr}  \{\mathcal{A}_r \, \textrm{has $l$ users},  \nonumber \\
&\qquad \, \mathcal{A}\,{-}\,\mathcal{A}_r \, \textrm{has at least $(i-l)$ users}\}+\sum \limits_{l=i}^{\infty} {\rm Pr}  \{\mathcal{A}_r \, \textrm{has $l$ users} \}, \nonumber \\
&  = \sum \limits_{l=k}^{i-1} \frac{e^{-\mathcal{A}_r\lambda}\left[ \mathcal{A}_r\lambda\right]^l}{l!}\left\lbrace 1 - \sum \limits_{l'=0}^{i-l-1} \frac{e^{-\frac{\Delta}{2}(L_2^2{-}r^2)\lambda}\left[\frac{\Delta}{2}(L_2^2{-}r^2)\lambda \right]^{l'}}{l'!}\right\rbrace \nonumber \\
&\hspace{12em}+ \left\lbrace 1 - \sum \limits_{l=0}^{i-1} \frac{e^{-\mathcal{A}_r\lambda}\left[ \mathcal{A}_r\lambda\right]^{l}}{l!}\right\rbrace \label{eq:CDF_K_greater_i_step_1}.
\end{align}\normalsize

% In addition, given that the number of users in $\mathcal{A}_k$ is $l\,{\geq}\,k$, the second condition $j \,{\leq}\, K \,{<}\, i$ requires that the remaining area $\mathcal{A}\,{-}\,\mathcal{A}_k$ has at most $(i\,{-}\,l\,{-}\,1)$ users, so that the user region has less than $i$ users. As a result, the desired probability is calculated as
% \begin{align}
% \hspace{-0.14in}{\rm P}\{ d_k \,{<}\, r_k, j \,{\leq}\, K \,{<}\, i \} &\,{=} \sum \limits_{l=k}^{i-1} {\rm P}  \{\mathcal{A}_k \, \textrm{has $l$ users}, \, \mathcal{A}{-}\mathcal{A}_k \, \textrm{has at most $(i\,{-}\,l\,{-}\,1)$ users}\} \nonumber \\
% &= \sum \limits_{l=k}^{i{-}1} \frac{e^{{-}\Delta(r_k^2{-}L_1^2)\lambda}\left[ \Delta(r_k^2{-}L_1^2)\lambda \right]^l}{l!}\left\lbrace \sum \limits_{l'=0}^{i{-}l{-}1} \frac{e^{{-}\Delta(L_2^2{-}r_k^2)\lambda}\left[ \Delta(L_2^2{-}r_k^2)\lambda \right]^{l'}}{l'!}\right\rbrace. \label{eq:marginal_cdf}
% \end{align}

\begin{figure*}[!htb]
\setcounter{MYtempeqncnt}{\value{equation}}
\setcounter{equation}{55} \footnotesize
\begin{align} \label{eq:PDF_ordered_abs_distance_Sk3}
f_{d_k|\mathcal{S}_{K_3}} (r)&= \frac{{\rm d}}{{\rm d} r } F_{d_k|\mathcal{S}_{K_3}} (r) = \frac{1}{\mathcal{C}} \frac{{\rm d}}{{\rm d} r}  \left\lbrace \sum \limits_{l=k}^{i-1} e^{-\mathcal{A}_r\lambda} \frac{\left[ \mathcal{A}_r\lambda \right]^l}{l!} \right\rbrace - \frac{e^{-\frac{\Delta}{2} L}}{\mathcal{C}}\frac{{\rm d}}{{\rm d} r}  \left\lbrace \sum \limits_{l=k}^{i-1} \sum \limits_{l'=0}^{i-l-1} \frac{\left[ \mathcal{A}_r\lambda\right]^l}{l!} \frac{\left[\frac{\Delta}{2}(L_2^2{-}r^2)\lambda \right]^{l'}}{l'!}\right\rbrace+\frac{1}{\mathcal{C}} \frac{{\rm d}}{{\rm d} r }  \left\lbrace 1- \sum \limits_{l=0}^{i-1} e^{-\mathcal{A}_r\lambda} \frac{\left[ \mathcal{A}_r\lambda \right]^l}{l!} \right\rbrace
\nonumber \\
&=  \frac{\Delta \lambda r}{\mathcal{C}} \left\lbrace e^{-\mathcal{A}_r\lambda} \left\lbrace  \frac{ \left[ \mathcal{A}_r\lambda \right]^{(k-1)}}{(k-1)!} - \frac{ \left[ \mathcal{A}_r\lambda \right]^{(i-1)}}{(i-1)!}  \right\rbrace -  \frac{ e^{-\frac{\Delta}{2} L} \left[ \mathcal{A}_r\lambda \right]^{(k-1)}}{(k-1)!} \left\lbrace \sum \limits_{l'=0}^{i-k-1} \frac{\left[ \frac{\Delta}{2}(L_2^2{-}r^2)\lambda \right]^{l'}}{l'!} \right\rbrace +  \frac{ e^{-\mathcal{A}_r\lambda } \left[ \mathcal{A}_r\lambda  \right]^{(i-1)}}{(i-1)!}  \right\rbrace
\nonumber \\
&=\frac{ \Delta \lambda r}{\mathcal{C}}\frac{\left[\mathcal{A}_r\lambda \right]^{(k-1)}}{(k-1)!} \left\lbrace e^{-\mathcal{A}_r\lambda}- e^{-\frac{\Delta}{2} L} \sum \limits_{l=0}^{i-k-1}\frac{\left[\frac{\Delta}{2}(L_2^2{-}r^2)\lambda \right]^{l}}{l!} \right\rbrace.
\end{align} \normalsize
\setcounter{equation}{56}
\hrulefill
\end{figure*}

Using \eqref{eq:CDF_K_greater_i_step_1} and $\mathcal{C}\,{=}\,1-\sum\limits_{l=0}^{i{-}1} \frac{e^{{-}\frac{\Delta}{2}L} \left[\frac{\Delta}{2}L \right]^l}{l!}$ with $L\,{=}\,(L_2^2 - L_1^2)\lambda$, the CDF in \eqref{eq:CDF_K_greater_i} can be readily obtained. Taking the derivative of \eqref{eq:CDF_K_greater_i} with respect to $r$, PDF can be derived as in \eqref{eq:PDF_ordered_abs_distance_Sk3}. \hfill \IEEEQEDhere %\vspace{-1em}

\bibliographystyle{IEEEtran}
\bibliography{references}
\end{document}